# Dissipative Coupling in Photonic and Plasmonic Resonators

Tong Wu[1] and Philippe Lalanne[1]*

[1]LP2N, CNRS, IOGS, Université Bordeaux, Talence, France

*philippe.lalanne@institutoptique.fr

**Abstract**: The rapid progress of nanophotonics demands theoretical frameworks capable of predicting the resonant behavior of complex systems comprising constituents of varying nature, operating beyond the weak-coupling, high-Q regime where classical temporal coupled-mode theory (CMT) is applicable. This work presents a coupled-quasinormal-mode (cQNM) framework for analyzing dissipative coupling with photonic and plasmonic resonators. The framework provides rigorous closed-form expressions for dissipative coupling coefficients and introduces novel features, such as a new coupling scheme via time derivatives of excitation coefficients. It delivers transparent and accurate predictions of exotic phenomena—such as zero-coupling between very close cavities and level-attraction effects—that are only vaguely captured by traditional CMT models. Efficient and user-friendly, this framework facilitates rapid parameter space exploration for device design and offers potential for extension to nonlinear and quantum systems in future applications.

## 1. Introduction

State hybridization usually refers to the rearrangement of the states of several individual bodies to create new states that are hybrids of the original states. It occurs across many areas of science, a famous example being the linear-combination-of-atomic-orbitals (LCAO) method in quantum chemistry. In the present context centered on nanophotonics, physics governing the coherent and dissipative couplings between distinct localized modes, e.g. plasmons, phonons, Mie resonances, or molecules bears a great resemblance to the electronic bonding of atoms [1]. The hybridization leads to the formation of new 'molecular' states with energies, decay rates, near-field properties and far field radiation patterns, which are completely different from the initial modes of the individual 'atoms' taken separately.

Early examples of hybridized states with spectacular impacts encompass various dimers formed with photonic [2] or plasmonic [3-5] resonances, chains and polymers [6,7] formed with many resonances, or hybrids [8,9] formed by the association of electromagnetic and material resonances. In recent years, coherent and dissipative couplings between resonances have enabled a wealth of interesting phenomena that can guide the design of novel devices in various areas, from cavity QED to nonlinear optics, vibrational spectroscopy, lasers and solid-state lighting devices. For instance, the hybridization of high-quality-factor optical mode and tightly confined plasmonic resonances have served as an outstanding platform for enhancing light-matter interaction [10-14]. Other fascinating phenomena include ultra-confined bright modes via plasmon coupling [15-17], parity-time symmetry breaking [18-20], intrinsic chiral resonant modes [21], subwavelength cavity modes with remarkably-large Q factor [22], nonlinear light generation with defect-protected topological states [23], anti-PT symmetric systems [20,24], level attraction through mode coupling [25], and a myriad of resonant nanoparticles with exotic scattering and absorbing properties [20,26-33].

In most studies in nanophotonics, the analysis of devices involving both coherent and dissipative couplings is typically performed with the (temporal) coupled mode theory (CMT) [34-38]. CMT operates analogously to the LCAO method, assuming that the coupled modes (the supermodes) can be represented by a linear combination of the initial uncoupled modes of the individual resonators.

For a pair of coupled resonators labeled as '$A$' and '$B$,' the key result of CMT is the evolution equation that governs the dynamics of the coupled resonators (with the $\exp(-i\omega t)$ convention)

$$i \begin{bmatrix} \dot{\boldsymbol{\beta}}^A \\ \dot{\boldsymbol{\beta}}^B \end{bmatrix} = \begin{bmatrix} \widetilde{\boldsymbol{\omega}}^A & \boldsymbol{\kappa} \\ \boldsymbol{\kappa}^T & \widetilde{\boldsymbol{\omega}}^B \end{bmatrix} \begin{bmatrix} \boldsymbol{\beta}^A \\ \boldsymbol{\beta}^B \end{bmatrix} + \begin{bmatrix} \mathbf{D}^A(t) \\ \mathbf{D}^B(t) \end{bmatrix}, \tag{1}$$

where $\boldsymbol{\beta}^A$ and $\boldsymbol{\beta}^B$ are the modal excitation coefficients of the uncoupled modes, $\dot{\boldsymbol{\beta}}$ denotes the time derivative of $\boldsymbol{\beta}$, $\widetilde{\boldsymbol{\omega}}^A$ and $\widetilde{\boldsymbol{\omega}}^B$ are the diagonal matrices of the complex resonance frequencies of the individual resonators, and $\mathbf{D}^A$ and $\mathbf{D}^B$ are the driving terms proportional to the incident electric field $\mathbf{E}_D$ applied to each resonator. The key parameter is the resonator-to-resonator coupling matrix $\boldsymbol{\kappa}$.

Equation (1) is valued for its intuitiveness, computational simplicity, and clear delineation of the underlying physics. It is commonly understood that in the absence of energy loss, the coupling between two resonators is mediated solely by the real part of the coupling coefficients $Re(\boldsymbol{\kappa})$, leading to Rabi oscillations and level repulsion. On the other hand, $Im(\boldsymbol{\kappa})$ is caused by the coupling through the continuum and is found in non-Hermitian systems [39-40]. $Re(\boldsymbol{\kappa})$ and $Im(\boldsymbol{\kappa})$ are often referred to as the coherent and dissipative coupling coefficients, respectively [41-45].

Equation (1) is widely used in micro- and nanophotonics. However, it should be regarded as a heuristic equation. The coupling matrix $\boldsymbol{\kappa}$ and the coefficients associated with the interaction with the incident field (implicitly contained in $\mathbf{D}^A$ and $\mathbf{D}^B$) are typically determined by fitting numerical or experimental data. The equation is built on two key assumptions [34,35]: (1) energy conservation in the coupled system, and (2) orthogonality of the two coupled modes. The second assumption implies that the coupling is proportional to the amplitude of the other resonator, rather than to its time derivative. These two assumptions result in symmetric coupling coefficients in Eq. (1).

However, it is important to note that for low-Q resonators, the concept of energy stored in a mode becomes ill-defined [45]. Additionally, when the modes of the two resonators are strongly coupled, they become highly non-orthogonal, further violating the foundational assumptions. Therefore, the equation is limited to systems with high-quality-factor modes [47-50] in the weak-coupling limit—two assumptions that do not hold in many emerging systems, such as dissipatively coupled resonators [14,31,40-43,45,51], plasmonic resonators with high intrinsic losses, and ultra-strongly coupled resonators [52,53], where coupling strengths may approach or exceed the characteristic mode frequencies.

In this work, we address these limitations and propose a rigorous, non-Hermitian framework for modeling resonator couplings. Building on recent advancements on the electromagnetic theory of quasinormal modes (QNMs) [54-65], we derive the evolution equation governing the dynamics of coupled resonators directly from Maxwell's equations, without approximations. The resulting equation,

$$i \begin{bmatrix} \dot{\boldsymbol{\beta}}^A \\ \dot{\boldsymbol{\beta}}^B \end{bmatrix} = \begin{bmatrix} \widetilde{\boldsymbol{\omega}}^A & \boldsymbol{\kappa}^{AB} \\ \boldsymbol{\kappa}^{BA} & \widetilde{\boldsymbol{\omega}}^B \end{bmatrix} \begin{bmatrix} \boldsymbol{\beta}^A \\ \boldsymbol{\beta}^B \end{bmatrix} - i \begin{bmatrix} 0 & \mathbf{g}^{AB} \\ \mathbf{g}^{BA} & 0 \end{bmatrix} \begin{bmatrix} \dot{\boldsymbol{\beta}}^A \\ \dot{\boldsymbol{\beta}}^B \end{bmatrix} + \begin{bmatrix} \mathbf{F}^A(t) \\ \mathbf{F}^B(t) \end{bmatrix}, \tag{2}$$

which we refer to as the coupled-QNM (cQNM) equation, provides highly accurate predictions for a wide range of systems. It also reveals key differences with the CMT equations, some of which are unexpected:

- Energy representation: In CMT, the squared magnitude of the modal excitation coefficient, i.e. $|\boldsymbol{\beta}|^2$, represents the energy stored in the mode. In the non-Hermitian context, the concept of stored energy becomes less meaningful and the $\beta$-coefficients follow a normalization scheme that does not rely on energy [66].
- Coupling matrices computation: The coupling matrix $\boldsymbol{\kappa}^{AB}$, $\boldsymbol{\kappa}^{BA}$, $\mathbf{g}^{AB}$, $\mathbf{g}^{BA}$, as well as the coupling to the incident fields — the driving term is from now denoted by $[\mathbf{F}^A, \mathbf{F}^B]$ to emphasize that it is different with that in Eq. (1) — have all simple closed-form expressions,

- which are essentially overlap integrals between different modes. No fitting is required like in Eq. (1).
- Time-derivative driving term: The driving term $[\mathbf{F}^A, \mathbf{F}^B]$, which in CMT is proportional to the incident electric field, now depends on both the electric field and its time derivative, echoing recent advancements in the evolution equation for single isolated resonators [67,68].
- Time-derivative coupling: Unlike the conventional CMT, where the coupling is solely based on the excitation coefficients $\boldsymbol{\beta}$, in the cQNM equation, the coupling also involves the time derivatives of these coefficients, $\dot{\boldsymbol{\beta}}$. In fact, the coupling is predominantly governed by $\dot{\boldsymbol{\beta}}$ for non-dispersive resonators, since as will be shown below, for non-dispersive materials, $\boldsymbol{\kappa}^{AB} = \boldsymbol{\kappa}^{BA} = \mathbf{0}$, implying that the coupling arises solely from the coefficients $\mathbf{g}^{AB}$ and $\mathbf{g}^{BA}$.

Time-derivative couplings also arise within the CMT formalism when considering the non-orthogonality of the coupled modes [34]. However, non-orthogonality is not the physical origin of the time-derivative coupling in the cQNM formalism. This distinction becomes evident when considering the case $\varepsilon_b = \varepsilon_\infty$, where time-derivative couplings vanish, even though the QNMs of the two resonators remain non-orthogonal. In our formalism, time-derivative couplings rather reflect the possible immediate response of the QNMs in one resonator to the excitation of QNMs in the other. This behavior is encountered in the driving-force term, which is proportional to the time derivative of the incident field, indicating a prompt QNM response to external excitation.

Equation (2) also exhibits marked differences with earlier studies on dissipative coupling with QNMs. For example, the coupling through time derivatives (the second term on the right-hand side of Eq. (2)) is a unique feature not found in earlier studies [69-76]. Additionally, the closed-form expressions for the coupling coefficients introduced here are entirely new. Compared to advanced existing frameworks for dispersive materials [71], the cQNM equation offers the significant advantage of yielding a linear eigenvalue problem, which greatly simplifies the determination of important outcomes, such as the supermodes of the coupled resonators.

The article is organized as follows. Section 2 introduces the cQNM framework, detailing all the intermediate theoretical steps leading to Eq. (2) and providing the necessary foundation for practical application of the framework.

In Section 3, numerical tests are presented to validate the cQNM framework, focusing on particularly challenging cases involving dispersive resonators with strong interactions, weak spatial confinement, and significant leakage (quality factors $Q \sim 5$).

Section 4 highlights the ability of the cQNM framework to interpret previously inaccessible coupling phenomena and demonstrates its practical utility in inverse design for nanophotonics—an area where brute-force computation alone is inadequate, and theoretical insight is crucial [77,78].

We conclude in Section 5.

The initial motivation for this work was to gain a clearer understanding of dissipative coupling in electromagnetism. While electromagnetic simulators have made tremendous progress and are increasingly widely used, we believe that the nanophotonic community has yet to fully capitalize on their potential. This is the main point we aim to illustrate in Section 4. We want to emphasize that this analysis is not intended to diminish the value of the CMT or its predictive capabilities. Rather, we hope that this additional perspective can contribute to its development, especially given the growing popularity of QNM open-source software.

## 2. Theory

We begin by introducing our notation. Consider two coupled resonators, $A$ and $B$, with a straightforward extension to multiple resonators (Suppl. Section S1). We adopt the usual scattering-field formulation, see Annex 2 in [45], which expresses the system permittivity $\varepsilon(\mathbf{r}, \omega) = \varepsilon_b(\mathbf{r}, \omega) +$

$\Delta\varepsilon^A(\mathbf{r},\omega) + \Delta\varepsilon^B(\mathbf{r},\omega)$, where $\varepsilon_b$ is the background permittivity and $\Delta\varepsilon^X$ (with $X \in \{A, B\}$) represents the permittivity change introduced by each resonator. The coupled system is illuminated by a monochromatic driving wave with an electric field $\mathbf{E}_D(\mathbf{r},\omega)$.

To ensure analytical prolongation in the complex frequency plane, we assume that the permittivities of all individual constituents follow a standard $N$-pole Lorentz-Drude model, $\varepsilon(\omega) = \varepsilon_\infty - \varepsilon_\infty \sum_n \frac{\omega_{p,n}^2}{\omega^2 + i\omega\gamma_n - \omega_{0,n}^2}$, where $\omega_{p,n}$ are the plasma frequencies, $\gamma_n$ the damping coefficients, and $\omega_{0,n}$ the resonant frequencies. This assumption holds for most materials at optical frequencies [79], though alternative permittivity models exist [80].

We use a regularized-QNM framework [81] hereafter and its companion **MAN** freeware [82] to compute the modes of the individual resonator used to expand the field in space. The framework relies on a rotation into the complex-frequency plane with perfectly matched layers (PMLs), which conveniently map the exponentially divergent QNMs in open space into square-integrable wavefunctions. In the mapped space (Fig. 1), the individual modes are source free solutions of Maxwell equations with a permitivity $\varepsilon_b + \Delta\varepsilon^X$. These modes encompass a restricted set of dominant QNMs and additional PML modes [57,63,81]. The dominant QNMs preserves physical insight and the PML modes mathematically guarantee the completeness of the modal expansion over the entire space, $\Omega = \Omega_{phy} + \Omega_{PML}$. The electromagnetic fields of the QNMs and PML modes are denoted by $[\tilde{\mathbf{E}}_m^X, \tilde{\mathbf{H}}_m^X]$, $m = 1,2,\ldots$, and are normalized such that $\iiint_\Omega \left[\tilde{\mathbf{E}}_m^X \cdot \frac{\partial(\omega\varepsilon)}{\partial\omega}\tilde{\mathbf{E}}_m^X - \tilde{\mathbf{H}}_m^X \cdot \frac{\partial(\omega\mu)}{\partial\omega}\tilde{\mathbf{H}}_m^X\right]d^3\mathbf{r} = 1$ [63,81].

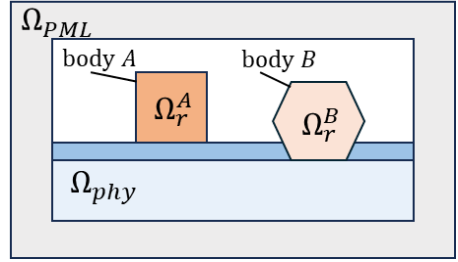

**Figure 1. Overview of the cQNM framework and notations.** The coupled system is composed of two nearby resonators, labelled body $A$ or $B$, each characterized by a permittivity change, $\Delta\varepsilon^A(\mathbf{r},\omega)$ or $\Delta\varepsilon^B(\mathbf{r},\omega)$, relatively to the background permittivity $\varepsilon_b(\mathbf{r},\omega)$. After applying a complex coordinate transformation for regularization, the mapped space is divided into an exterior perfectly matched layer (PML) domain ($\Omega_{PML}$), shown in grey, and the inner physical domain ($\Omega_{phy}$), which encompasses the resonators.

In the expanded basis comprising QNMs and PML modes, completeness is guaranteed, allowing us to express the electric fields $\mathbf{E}_S^A$ scattered by resonator $A$ as a superposition of QNMs and PML modes

$$\mathbf{E}_s^A(\mathbf{r},\omega) = \sum_m \alpha_m^A(\omega)\tilde{\mathbf{E}}_m^A(\mathbf{r}), \tag{3}$$

for any $\mathbf{r} \in \Omega$. A similar expression holds for resonator $B$. In Eq. (3), $\alpha_m^A$ is the modal excitation coefficient of the mode $m$ of resonator $A$. The excitation is due to the total field driving resonator $A$, i.e. the sum of the incident field $\mathbf{E}_D$ and the field $\mathbf{E}_S^B$ scattered by resonator $B$ on $A$. It is given by [81]

$$\alpha_m^A(\omega) = \iiint_{\Omega_r^A} \left(\frac{\tilde{\omega}_m^A}{\tilde{\omega}_m^A - \omega}\Delta\varepsilon^A(\tilde{\omega}_m^A) + [\varepsilon_b - \varepsilon_\infty^A]\right)\tilde{\mathbf{E}}_m^A(\mathbf{r}) \cdot \left(\mathbf{E}_D(\mathbf{r},\omega) + \mathbf{E}_s^B(\mathbf{r},\omega)\right)d^3\mathbf{r}. \tag{4}$$

Similar equations are written in an equivalent manner for resonator $B$. Equation (3) and its sister equation obtained by swapping $A$ and $B$ then generate a system of coupled equations in which each resonator is excited by the incident field and by the field scattered by the other resonator. Denoting the vector formed by all modal excitation coefficients by $|\boldsymbol{\alpha}\rangle = [\alpha_1^A, \alpha_2^A \ldots \alpha_1^B, \alpha_2^B \ldots]^T$, the coupled system can be written as

$$\widehat{\mathbf{H}}|\boldsymbol{\alpha}(\omega)\rangle = \omega\widehat{\mathbf{D}}|\boldsymbol{\alpha}(\omega)\rangle + |\mathbf{F}(\omega)\rangle, \tag{5}$$

with $\widehat{\mathbf{H}} = \begin{bmatrix} \widetilde{\boldsymbol{\omega}}^A & \boldsymbol{\kappa}^{AB} \\ \boldsymbol{\kappa}^{BA} & \widetilde{\boldsymbol{\omega}}^B \end{bmatrix}$ and $\widehat{\mathbf{D}} = \begin{bmatrix} \mathbf{I} & \mathbf{g}^{AB} \\ \mathbf{g}^{BA} & \mathbf{I} \end{bmatrix}$, where $\mathbf{I}$ is the identity the matrix.

The driving term in Eq. 5 is proportional to the monochromatic incident field $\mathbf{E}_D$. It is defined by $|\mathbf{F}(\omega)\rangle = \left[\alpha_{1,D}^A(\widetilde{\omega}_1^A - \omega), \alpha_{2,D}^A(\widetilde{\omega}_2^A - \omega) \ldots \alpha_{1,D}^B(\widetilde{\omega}_1^B - \omega), \alpha_{2,D}^B(\widetilde{\omega}_2^B - \omega)\right]^T$, with $\alpha_{m,D}^X$ ($X \equiv A$ or $B$) given by Eq. (4) with $\mathbf{E}_s^X \equiv \mathbf{0}$. In practice, once the QNMs of each individual resonator are determined, the matrices $\widehat{\mathbf{H}}$ and $\widehat{\mathbf{D}}$ along with the vector $|\mathbf{F}(\omega)\rangle$ can be readily computed. The fields scattered by the resonators, as described by Eq. (2), are then efficiently obtained via matrix inversion: $|\boldsymbol{\alpha}(\omega)\rangle = \left[\widehat{\mathbf{H}} - \omega\widehat{\mathbf{D}}\right]^{-1}|\mathbf{F}(\omega)\rangle$ [71]. Finally, the total field is obtained by summing the scattered fields of the resonators and the incident field: $\mathbf{E}_{tot}(\mathbf{r},\omega) = \mathbf{E}_s^A(\mathbf{r},\omega) + \mathbf{E}_s^B(\mathbf{r},\omega) + \mathbf{E}_D(\mathbf{r},\omega)$.

Equation (5) can be converted into the time domain using Fourier transforms to obtain the previously announced evolution Eq. (2), where $\beta^X(t) = \int_{-\infty}^{\infty} \alpha^X(\omega) \exp(-i\omega t)\, d\omega$ and

$$\kappa_{mn}^{AB} = -\widetilde{\omega}_m^A \iiint_{\Omega_r^A} \varepsilon_L^A(\widetilde{\omega}_m^A) \widetilde{\mathbf{E}}_m^A \cdot \widetilde{\mathbf{E}}_n^B d^3\mathbf{r}, \tag{6a}$$

$$g_{mn}^{AB} = \iiint_{\Omega_r^A} \Delta\varepsilon_\infty^A \widetilde{\mathbf{E}}_m^A \cdot \widetilde{\mathbf{E}}_n^B d^3\mathbf{r}, \tag{6b}$$

$$F_m^A(t) = \iiint_{\Omega_r^A} \widetilde{\mathbf{E}}_m^A \cdot \left[\widetilde{\omega}_m^A \varepsilon_L^A(\widetilde{\omega}_m^A)\, \mathbf{E}_D(\mathbf{r},t) + i\Delta\varepsilon_\infty^A\, \dot{\mathbf{E}}_D(\mathbf{r},t)\right] d^3\mathbf{r}. \tag{6c}$$

with $\Delta\varepsilon_\infty^X = \varepsilon_\infty^X - \varepsilon_b$ and $\varepsilon_L^X(\omega) = \varepsilon^X(\omega) - \varepsilon_\infty^X$. Similar expressions hold for $\kappa^{BA}$, $g^{BA}$ and $F_m^B(t)$.

Equation (6) stands as the pivotal outcome of the cQNM theory and constitutes our first important result, providing a precise definition of all coupling coefficients between the modes and the incident field. Equations (2) and (5) remain rigorous and free from restrictive assumptions, maintaining accuracy even for highly dispersive and lossy materials. The main limitation arises when the separation distance between the resonators is large [83]. In this regime, interactions are predominantly mediated by leaky (divergent) waves rather than evanescent waves, requiring the inclusion of numerous PML modes in the expansion to maintain accuracy. However, this compromises physical intuition, ultimately rendering the formalism impractical. Fortunately, such large separation distances are rarely relevant in nanophotonics [83].

So far, we have discussed the response of the coupled resonators in terms of the individual modes. However, it may be more insightful to represent this response in the supermode basis, i.e., in terms of the natural resonances of the system. The supermodes are denoted by $|\boldsymbol{\alpha}_m^S\rangle$ and their eigenfrequencies by $\widetilde{\omega}_m^S$ (the superscript 'S' stands for 'super'). They can be determined by solving the eigenproblem,

$$\widehat{\mathbf{H}}|\boldsymbol{\alpha}_m^S\rangle = \widetilde{\omega}_m^S\, \widehat{\mathbf{D}}|\boldsymbol{\alpha}_m^S\rangle. \tag{7}$$

Since both $\widehat{\mathbf{H}}$ and $\widehat{\mathbf{D}}$ are independent of the frequency, the eigenproblem is linear and the state vector $|\boldsymbol{\alpha}(\omega)\rangle$ can be conveniently projected onto the supermode basis via a simple change of basis, $|\boldsymbol{\alpha}(\omega)\rangle = \sum_m c_m^S(\omega)|\boldsymbol{\alpha}_m^S\rangle$, with

$$c_m^S(\omega) = \frac{1}{(\widetilde{\omega}_m^S - \omega)} \frac{\langle \boldsymbol{\alpha}_m^S|\mathbf{F}\rangle}{\langle \boldsymbol{\alpha}_m^S|\widehat{\mathbf{D}}|\boldsymbol{\alpha}_m^S\rangle}, \tag{8}$$

where $\langle \boldsymbol{\alpha}_m^S|$ denotes the left eigenvector and satisfies the equation $\langle \boldsymbol{\alpha}_m^S|\widehat{\mathbf{H}} = \widetilde{\omega}_m^S \langle \boldsymbol{\alpha}_m^S|\widehat{\mathbf{D}}$. Note that, as $\widehat{\mathbf{H}} \neq \widehat{\mathbf{H}}^T$ and $\widehat{\mathbf{D}} \neq \widehat{\mathbf{D}}^T$ in general, $\langle \boldsymbol{\alpha}_m^S|$ and $|\boldsymbol{\alpha}_m^S\rangle^T$ are not colinear and $\langle \boldsymbol{\alpha}_m^S|$ requires an additional computation.

Equation (8) provides a closed-form expression of the supermode excitation coefficients and constitutes the second major result of this work. This closed-form expression is a unique feature of our formalism, which distinguishes it from previous works on cQNM theory with dispersive resonators [71,72]. In contrast to previous advanced theory [71], it leads to a linear eigenproblem due to the adoption of auxiliary fields in the cQNM framework. The linearity offers significant advantages in computing and manipulating the supermodes with simple linear matrix algebra.

Another unique aspect of our approach, not found in previous works, is the normalization of the supermodes. Supplementary Section S2 outlines two rigorous methods and Suppl. Section S3 also includes an approximate method, which preserves full analyticity in the supermode basis, enabling closed-form expressions for important physical quantities. QNM normalization holds intrinsic physical significance: the field intensity of the normalized QNMs tells the QNM capability to response to the driving field; the phase encodes the non-Hermitian character of QNMs and determines the line shape of the Purcell factor; the inverse of the square of the normalized QNM corresponds to the complex mode volume [66]. Moreover, the approximate method expands the capabilities of the current cQNM framework, enabling a detailed analysis of how the coupling of individual resonators affects the intrinsic properties of the supermodes in the coupled system. An illustrative example is provided in Section 4.1.

## 3. Numerical convergence test

In practice, we determine the QNMs and PML modes of the individual resonators using the mode solver QNMEig from the **MAN** freeware [82]. This solver automatically finds a large number of modes near a user-defined complex frequency. Since the QNMs and PML modes are treated equivalently within the cQNM framework, their coupling coefficients and driving-force terms follow the same closed-form equations.

Next, we truncate the expansion bases, which requires systematically sorting the modes by significance. To achieve this, we employ the sorting strategy in **MAN**, which ranks modes based on their excitation probability under plane wave illumination. This approach effectively differentiates QNMs from PML modes, identifies the dominant QNMs, and orders them accordingly (see the tutorial in [82] for detail).

Let $N_A$ and $N_B$ denote the truncation ranks for the QNM expansion of resonators $A$ and $B$, as defined in Eq. (3). Once the modes are sorted, we compute all coupling coefficients in Eq. (6), yielding the matrices $\widehat{\mathbf{H}}$ and $\widehat{\mathbf{D}}$ of size $(N_A + N_B) \times (N_A + N_B)$ and the driving vector $|\mathbf{F}(\omega)\rangle$ of $(N_A + N_B) \times 1$. We then solve Eqs. (2) and (5) either iteratively in the time domain or via matrix inversion in the frequency domain. These calculations are straightforward, involving overlap integral computations, matrix diagonalizations, and inversions of low-dimensional matrices.

For the test, a geometry consisting of two parallel and identical silicon wires of infinite length is considered (Fig. 2a). It incorporates a realistic, frequency-dependent permittivity for silicon, modeled using a single-pole Lorentz model (see caption). This approach allows us to test the formalism in its most general form, where all coupling coefficients, $\boldsymbol{\kappa}^{AB}$, $\boldsymbol{\kappa}^{BA}$, $\mathbf{g}^{AB}$, $\mathbf{g}^{BA}$, are non-zero. As expected, these coefficients are all complex, with significant imaginary parts, indicating that dissipative coupling plays a crucial role in this highly non-Hermitian system (Fig. 2b). Additionally, due to the strong dispersion of silicon in the visible spectral range, the coupling coefficients $\boldsymbol{\kappa}^{AB}$ and $\boldsymbol{\kappa}^{BA}$ are far from negligible, and are comparable to $\mathbf{g}^{AB}$ and $\mathbf{g}^{BA}$ when normalized by the central frequency of the concernted spectrum $\omega_0 = 2\pi c/(600 \text{ nm})$.

At first glance, the wire dimer geometry may seem simple, but it is, in fact, quite challenging. At visible frequencies, the dominant QNMs of the wires exhibit weak spatial confinement, along with significant leakage and absorption, with most having quality factors $Q < 5$. Consequently, accurately reconstructing the field scattered by a single wire illuminated by a monochromatic plane wave requires many modes. The challenge intensifies when the wires are paired. Due to electromagnetic interactions, the QNMs strongly couple not only with each other but also with PML modes, even for small separation distances.

Figure 2c reports on convergence tests conducted for a few supermodes of the nanowire dimer. We compare the frequency $\widetilde{\omega}_n^S(\text{cQNM})$, computed by solving Eq. (7), with the frequency $\widetilde{\omega}_n^S(\text{exact})$, computed directly for the dimer using **MAN** freeware [82]. The relative frequency error is defined as $|\widetilde{\omega}_n^S(\text{exact}) - \widetilde{\omega}_n^S(\text{cQNM})|/|\widetilde{\omega}_n^S(\text{exact})|$. Contrasted behaviors are observed. For high-$Q$ supermodes, $Q \approx 500$, fast convergence is achieved; only one mode for each resonator is needed to achieve a relative error smaller than $10^{-3}$. For low-$Q$ supermodes, $Q \approx 5$ and 14, the convergence is significantly slower: typically, 100 modes are required to reach a relative accuracy of $10^{-2}$. The hybridization is influenced not only by the monopolar and dipolar QNMs (with $Q \approx 1$ and 7, respectively) of the isolated nanowires but also by plenty of other QNMs and PML modes. Note that the frequencies of the monopolar and dipolar individual QNMs are $123 - 55i$ THz and $446 - 31i$ THz, respectively.

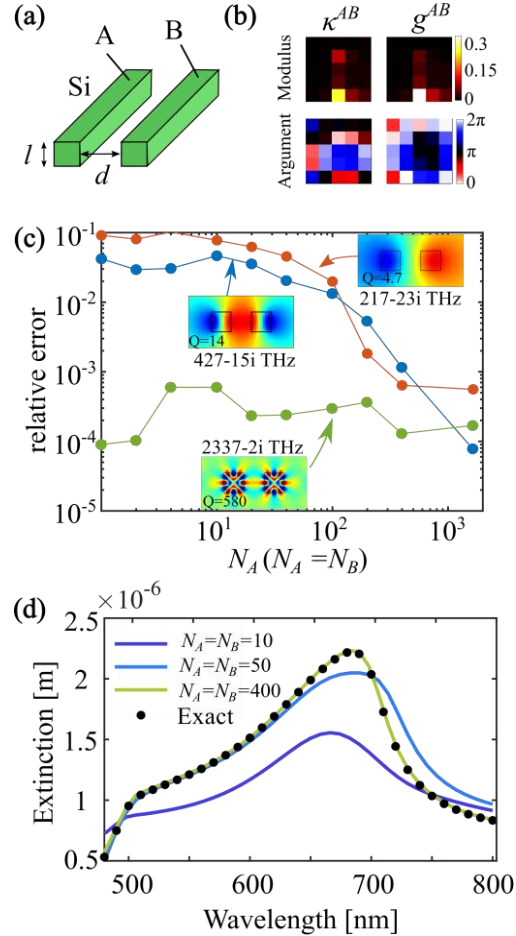

**Figure 2. Benchmarking the cQNM theory on a challenging example involving strong coupling between weakly confined low-Q modes.** (**a**) The geometry consists of two identical Si-nanowires with side length $l = 114$ nm and separation distance $d = 116$ nm. The frequency-dependent permittivity of silicon is accurately given by a single-pole Lorentz model with $\varepsilon_\infty = 8.51\varepsilon_0$, $\omega_{p,1} = 3.62 \times 10^{15}$ rad.s$^{-1}$, $\omega_{0,1} = 5.09 \times 10^{15}$ rad.s$^{-1}$, and $\gamma_1 = 1.16 \times 10^{14}$ rad.s$^{-1}$ in the 400–800 nm spectral range. (**b**) Coupling coefficients for five dominant modes (see Suppl. Section S4 for their field distributions). Only $\boldsymbol{\kappa}^{AB}$ and $\mathbf{g}^{AB}$ are shown in the figure. For symmetry reasons, $\boldsymbol{\kappa}^{AB} = \boldsymbol{\kappa}^{BA}$ and $\mathbf{g}^{AB} = \mathbf{g}^{BA}$. Absolute values and phases are provided in the first and second rows, respectively. The absolute value of $\boldsymbol{\kappa}^{AB}$ is normalized by $\omega_0 = 2\pi c/(600 \text{ nm})$ to compare with $\mathbf{g}^{AB}$. (**c**) Eigenfrequency convergence test of Eq. (7) for three superQNMs as $N_A = N_B$ increases. (**d**) Convergence test of Eq. (8). The extinction cross-section spectra computed with the cQNM theory for $N_A = N_B = 10, 50, 400$ supermodes are compared with COMSOL real-frequency simulation data ('Exact'). All computations are carried out for electric fields parallel to the wires.

Figure 2d tests the ability to accurately reconstruct the scattered field in the supermode basis with Eqs. (7) and (8). The extinction cross-section spectra of the dimer, illuminated by a normally incident plane wave, are compared with COMSOL real-frequency simulation data ('Exact'). Even for a relatively small number of modes ($N_A = N_B = 10$), the asymmetric line shape of the extinction spectrum is reasonably well predicted, although the amplitude is approximately 1.5 times smaller than the exact value. As $N_A$ increases up to 50, the discrepancy decreases, and excellent agreement is achieved for $N_A = 400$. This agreement, along with the very small values $\sim 10^{-4} - 10^{-5}$ of the relative error in Fig. 2c, clearly demonstrates the cQNM-theory capability to achieve high accuracy, even for challenging cases.

However, using a large number of QNMs and PML modes is computationally demanding and offers limited physical insight. It is important to reemphasize that we deliberately selected a complex scenario to rigorously test convergence. In contrast, CMT might still yield qualitatively good predictions of the extinction spectra by fitting coupling and frequency coefficients, even is the spectral shape appears unusually complex. From this perspective, CMT can appear more effective and insightful. However, one may question the relevance of such modal approaches: what do we truly learn from a modal description that does not reflect the actual modal structure of the system?

Illustrative examples of this concern can be found in [55] on the Purcell factor of a plasmonic dimer, and in [83] on the textbook case of two resonant polarizabilities. Section 4.3 further expands the discussion. Perhaps we must accept that coupled mode theories are not universally applicable with the current state of the art—and should be used only when they genuinely provide insight.

The convergence test reinforces the legitimacy of our cQNM framework. However, the true strength of the approach lies in its ability to provide quantitative insight and computational efficiency while retaining only a few dominant QNMs. An illustrative demonstration of this is presented in Suppl. Section S5, where we analyze the association of metallic nanorods supporting highly tunable, polarization-sensitive longitudinal plasmon modes. We show that the cQNM framework enables efficient prediction of the Fano resonance line shapes and wavelength for various 'Lego' plasmonic hybrids – incorporating a range of positions, orientations, and configurations for the individual resonators – by working with matrices typically smaller than 10×10. This makes initial designs based on the exploration of large parameter spaces computationally feasible.

In general, the cQNM theory is well-suited for studying the coupling between resonators that exhibit strong photon confinement, such as plasmonic antennas and photonic microcavities. In these cases, accurate results can typically be obtained by considering only a few QNMs, without the need to account for PML modes, as discussed in the following Section.

## 4. Applications

Tailoring mode hybridization to create new supermodes with properties largely different from the original modes is a crucial step in inverse design in nanophotonics [4,5,78]. This process typically involves repeatedly solving Maxwell's equations and then using the CMT with fitted coefficients to interpret the operation of the optimized device. However, this approach fails when non-Hermitian couplings involving substantial absorption or leakage into a continuum result in unexpected phenomena [14,19,44,51].

In this section, we highlight the unique advantages of the cQNM framework with three illustrative examples of unconventional coupling scenarios. The first revisits the fundamentals of electromagnetic mode hybridization, examining the impact of hybridization on the main resonator parameters, i.e. the quality factor and mode volume. The second demonstrates how the cQNM framework can inspire innovative design strategies, while the third shows its ability to provide clear insights into complex physical phenomena that were previously interpreted in a convoluted manner.

## 4.1 Engineering Q and V through hybridization: critical case of exceptional points

To explore the potential of mode hybridization, we analyze its impact on the three fundamental parameters of resonators: resonant frequency, quality factor, and mode volume.

Hermitian modes are stationary, with strictly real field distributions, resulting in real and positive mode volumes [66]. For most non-Hermitian resonators with quality factors $Q > 5$, the real part of the QNM field typically dominates over the imaginary part, yielding mode volumes with a predominantly positive real component. However, in coupled systems, special cases can arise where, despite the constituent resonators having nearly real mode volumes, the resulting supermodes represent genuinely non-Hermitian states. These may exhibit mode volumes with large imaginary components or even significantly negative real parts. In this section, we use the present theoretical framework to explain the emergence of such complex mode volumes in coupled systems.

We keep the discussion at a general level without specifying a particular system. We consider two resonators, $A$ and $B$, with QNM frequencies $\widetilde{\omega}^A$ and $\widetilde{\omega}^B$ and normalized QNM electric fields $\tilde{\mathbf{E}}^A$ and $\tilde{\mathbf{E}}^B$. The superQNMs of the system are found by solving Eq. (7), which becomes a 2×2 generalized eigenproblem $\begin{bmatrix} \widetilde{\omega}^A & \kappa^{AB} \\ \kappa^{BA} & \widetilde{\omega}^B \end{bmatrix} \begin{bmatrix} \tilde{\alpha}^A_\pm \\ \tilde{\alpha}^B_\pm \end{bmatrix} = \widetilde{\omega}_\pm \begin{bmatrix} 1 & g^{AB} \\ g^{BA} & 1 \end{bmatrix} \begin{bmatrix} \tilde{\alpha}^A_\pm \\ \tilde{\alpha}^B_\pm \end{bmatrix}$ with our single-mode approximation. The eigenfrequencies $\widetilde{\omega}_\pm$ and corresponding eigenvectors $[\tilde{\alpha}^A_\pm, \tilde{\alpha}^B_\pm]^T$ of the superQNMs $\tilde{\mathbf{E}}_\pm(\mathbf{r}) = \tilde{\alpha}^A_\pm \tilde{\mathbf{E}}^A(\mathbf{r}) + \tilde{\alpha}^B_\pm \tilde{\mathbf{E}}^B(\mathbf{r})$ are then easily computed. To achieve full analyticity, we assume that the couplings $\widetilde{\omega}_\pm g^{AB}, \widetilde{\omega}_\pm g^{BA}, \kappa^{AB}, \kappa^{BA}$ are small in norm compared to $\widetilde{\omega}^{A(B)}$, and the eigenfrequencies of the supermodes are close to half of the sum of the frequencies of the uncoupled resonators, $\Omega = \frac{\widetilde{\omega}^A + \widetilde{\omega}^B}{2}$. Then by introducing the effective coupling parameters, $\chi^{AB} = \kappa^{AB} - \Omega g^{AB}$ and $\chi^{BA} = \kappa^{BA} - \Omega g^{BA}$, we approximate the generalized eigenvalue problem by a standard form

$$\begin{bmatrix} \widetilde{\omega}^A & \chi^{AB} \\ \chi^{BA} & \widetilde{\omega}^B \end{bmatrix} \begin{bmatrix} \tilde{\alpha}^A_\pm \\ \tilde{\alpha}^B_\pm \end{bmatrix} = \widetilde{\omega}_\pm \begin{bmatrix} \tilde{\alpha}^A_\pm \\ \tilde{\alpha}^B_\pm \end{bmatrix}. \tag{9}$$

Solving this eigenvalue problem yields

$$\widetilde{\omega}_\pm = \Omega \pm \sqrt{\Omega^2 + \chi^{BA}\chi^{AB} - \widetilde{\omega}^A \widetilde{\omega}^B}. \tag{10}$$

Admittedly, the cQNM formula yields a similar prediction of the eigenfrequencies compared to the conventional CMT equation. However, as we will show below, the present theory provides an explicit expression for the mode volume, thereby elucidating the origin of complex mode volumes in coupled systems.

As shown in Suppl. Section S3, under the condition $\widetilde{\omega}^A \approx \widetilde{\omega}^B \approx \widetilde{\omega}_\pm$, the supermode normalization simplifies to $(\tilde{\alpha}^A_\pm)^2 + (\tilde{\alpha}^B_\pm)^2 = 1$, and the coupling coefficients can be approximated as $\chi^{BA} = \chi^{AB}$. Using these approximations, we obtain: $(\tilde{\alpha}^A_\pm)^2 = \frac{1}{2}\frac{(\Delta \pm \sqrt{R})^2}{R \pm \Delta\sqrt{R}}$ and $(\tilde{\alpha}^B_\pm)^2 = (\tilde{\alpha}^A_\mp)^2$, where $R = \Delta^2 + 4\chi^2$, $\Delta = \widetilde{\omega}^A - \widetilde{\omega}^B$, and $\chi = \frac{1}{2}(\chi^{AB} + \chi^{BA})$. Finally, using the normalized superQNM and the definition of mode volume, we obtain the closed-form expression

$$\frac{1}{\tilde{V}^S_\pm} = \frac{(\tilde{\alpha}^A_\pm)^2}{\tilde{V}^A} + \frac{(\tilde{\alpha}^B_\pm)^2}{\tilde{V}^B} + 2\tilde{\alpha}^A_\pm \tilde{\alpha}^B_\pm \sqrt{\frac{1}{\tilde{V}^A}\frac{1}{\tilde{V}^B}}, \tag{11}$$

which expresses the mode volumes $\tilde{V}^S_\pm$ of the superQNMs in terms of the mode volumes, $\tilde{V}^A$ and $\tilde{V}^B$, of the isolated resonators. $\tilde{V}^A$ and $\tilde{V}^B$ are given by $(\tilde{V}^A)^{-1} = 2\varepsilon_0 (\mathbf{u} \cdot \tilde{\mathbf{E}}^A)^2$ and $(\tilde{V}^B)^{-1} = 2\varepsilon_0 (\mathbf{u} \cdot \tilde{\mathbf{E}}^B)^2$, where $\mathbf{u}$ is a unit vector aligned with the polarization direction of a small object [66]. Similarly, $(\tilde{V}^S_\pm)^{-1} = 2\varepsilon_0 (\mathbf{u} \cdot \tilde{\mathbf{E}}_\pm)^2$.

The expressions in Eqs. (10-11) are derived under certain approximations, yet they maintain considerable generality. They are applicable to any dimers, including coupled high-Q photonic microcavities, or hybrid dimers combining high-Q cavities with small-V plasmonic antennae.

To evaluate their precision, we examine a gap antenna composed of two closely positioned gold nanorods. We first calculate the supermodes directly using **MAN**. This reveals two dominant superQNMs around 630 nm. In this wavelength range, the normalized spontaneous emission rate of a source located at the center of the gap and polarized along the z-axis exceeds 100 [55]. The "exact" mode volumes of the superQNMs are shown in Fig. 3 and are compared with those predicted by Eq. (11), using the individual dipolar-like modes of each nanorod with complex wavelengths $2\pi c/\tilde{\omega}^A = 652 + 20i$ nm and $2\pi c/\tilde{\omega}^B = 620 + 98i$ nm. A quantitative agreement is achieved. The observed discrepancies at certain locations primarily stem from limitations of the two-mode approximation.

Equation (11) provides insight into how coupling transforms quasi-Hermitian states into non-Hermitian states. The transformation is governed by the eigenvector coefficients, $\tilde{\alpha}_\pm^A$ and $\tilde{\alpha}_\pm^B$, which are complex numbers in general.

This is illustrated in Fig. 3, where the mode volumes $\tilde{V}^A$ and $\tilde{V}^B$ of the individual resonators are predominantly real and positive. However, despite $\tilde{V}^A$ and $\tilde{V}^B$ being nearly real, the resulting supermodes exhibit mode volumes with either large imaginary components or significantly negative real values, depending on the spatial position within the maps. This highlights that couplings fundamentally alter the Hermitian character of the modes.

To fully understand the conditions under which this transformation becomes significant, we consider two extreme cases: strong and weak coupling. In the strong coupling regime, where $|\chi| \gg |\Delta|$, it follows from Eq. (9) that $(\tilde{\alpha}_\pm^A)^2 \approx (\tilde{\alpha}_\pm^B)^2 \approx 1/2$. Since $\tilde{\mathbf{E}}^A$ and $\tilde{\mathbf{E}}^B$ are predominantly real, the superQNM mode volumes exhibit very low values of $\text{Im}(\tilde{V}_\pm^S)/\text{Re}(\tilde{V}_\pm^S)$. Conversely, in the weak coupling regime, where $|\chi| \ll |\Delta|$, the superQNM mode volumes resemble those of the uncoupled resonators, remaining largely real and positive.

The system in Fig. 3 represents a critical regime with $\Delta/\chi = -0.06 - 1.98i$. In this regime, where $|R| \ll |\Delta^2|$, the coefficients can be approximated as

$$(\tilde{\alpha}_\pm^A)^2 = (\tilde{\alpha}_\mp^B)^2 \approx \pm \tfrac{1}{2}\Delta R^{-1/2}. \tag{12}$$

Since both $\Delta$ and $R$ are complex valued with comparable real and imaginary parts, the mode volumes inevitably acquire large imaginary parts.

A closer examination of the eigenvector expression reveals that the coupled nanorods operate near an exceptional point condition, where $\Delta/\chi = \pm 2i$. Near exceptional points where both eigenvalues and eigenvectors coalesce, a fine tuning of the system parameters—such as $\Delta$— allows for the realization of mode volumes spanning a broad range of $\text{Im}(\tilde{V}_\pm^S)/\text{Re}(\tilde{V}_\pm^S)$ values.

Finally, it is important to note that since $(\tilde{\alpha}_+^A)^2 \approx -(\tilde{\alpha}_-^A)^2$ and $(\tilde{\alpha}_+^B)^2 \approx -(\tilde{\alpha}_-^B)^2$, the real and imaginary parts of $\tilde{V}_+^S$ and $\tilde{V}_-^S$ exhibit nearly opposite signs, as shown in Fig. 3. This aligns with previous studies [55,84,85], which have demonstrated that the local density of states often features Fano lineshapes that deviate significantly from Lorentzian profiles near exceptional points.

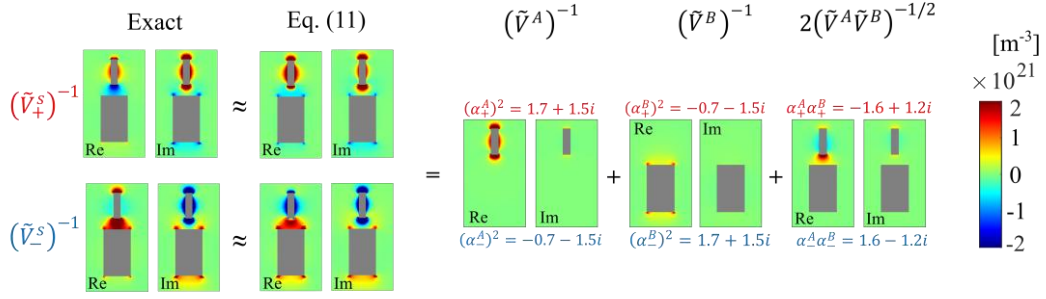

**Figure 3. Mode volume engineering through hybridization.** Two gold cylinders (radii $r_1 = 20$ nm, $r_2 = 42.5$ nm; lengths $L_1 = 80$ nm, $L_2 = 150$ nm) embedded in air are separated by a gap of $g = 38$ nm. The superQNMs calculated using **MAN** (Exact) are compared with the predictions of the closed-form expression of Eq. (11). The 3-term decomposition of Eq. (11) provided on the right-side of the figure highlights how, starting from real-valued mode volumes for the initial uncoupled QNMs, the hybridization produces negative or imaginary mode volumes, via the complex-valued coefficients $(\tilde{\alpha}_\pm^A)^2$, $(\tilde{\alpha}_\pm^B)^2$ and $\tilde{\alpha}_\pm^A \tilde{\alpha}_\pm^B$. The gold permittivity is modeled using a single-pole Drude approximation, $\varepsilon_\infty = \varepsilon_0$, $\omega_{p,1} = 1.26 \times 10^{16}$ Hz, $\gamma_1 = 0.011 \omega_{p,1}$ in the simulation.

### 4.2 Zero coupling between nearby cavities

For high-density photonic integration, minimizing coupling between nearby devices, such as waveguides or cavities, is crucial [86,87]. Previous studies on photonic crystal cavities have tackled this challenge using intuitive initial designs followed by iterative, blind numerical adjustments—without explicitly computing coupling coefficients or achieving a clear physical interpretation [87-89].

Here, we use the cQNM framework to develop an analytical method based on a perturbation lookup table. This approach provides direct guidance on modifying the photonic crystal layout to nullify both the real and imaginary parts of the coupling coefficient. Unlike earlier approaches, it offers transparent physical insights and novel design solutions that cannot be anticipated by intuition.

We consider a photonic dimer consisting of two identical photonic crystal cavities in a suspended semiconductor membrane (see Fig. 4a). Each cavity is formed by a missing hole and two shifted holes at the edges, separated by a single central hole. The holes are assumed to have vertical slopes. This ultra-compact layout, with minimal mode volume and separation distance, presents a rigorous test case for our method. The energy separation between the bonding and anti-bonding superQNMs is approximately ten times their linewidths, confirming strong coupling between the two cavities of the initial layout.

The cQNM framework is particularly straightforward for our system: since the cavities are identical, we have $g^{AB} = g^{BA} = g$, and because material dispersion is negligible in the narrow spectral range of interest, $\kappa^{AB} = \kappa^{BA} = 0$. The uncoupled cavities and their field distributions of $\tilde{\mathbf{E}}^A$ and $\tilde{\mathbf{E}}^B$ are shown in Suppl. Section S6.

A conventional approach using the cQNM framework would involve selecting a hole deformation, computing the QNMs of the isolated cavities with modified holes, accepting the deformation if the coupling coefficient $g$ (given by Eq. (6)) decreases, and iterating the process. However, this iterative procedure is computationally expensive. Instead, we bypass it entirely using perturbation theory, computing a single lookup table based solely on the QNM electric fields $\tilde{\mathbf{E}}^A$ and $\tilde{\mathbf{E}}^B$ of the isolated cavities.

For simplicity, we focus on varying the size of the holes. However, the formalism remains fully general and applies to arbitrary hole deformations, as illustrated in Fig. 4b. This figure defines the initial circular hole boundary ($\Sigma$) and the deformed boundary ($\Sigma'$), with the gray-shaded region representing the deformed area $\Delta A$. The deformation is parameterized by $h(\mathbf{r}_\Sigma)\hat{\mathbf{n}}$, where $\hat{\mathbf{n}}$ is the outward unit normal to $\Sigma$ and $\mathbf{r}_\Sigma$ is a coordinate along $\Sigma$. The function $h(\mathbf{r}_\Sigma)$ quantifies the perpendicular deformation from $\Sigma$ to $\Sigma'$, with $h > 0$ indicating outward deformation and $h < 0$ indicating inward deformation.

We apply a well-known first-order perturbation-theory result [90,91] to determine the QNM electric fields, $\tilde{\mathbf{E}}_D^A$ and $\tilde{\mathbf{E}}_D^B$, of the deformed cavities using only the initial fields $\tilde{\mathbf{E}}^A$ and $\tilde{\mathbf{E}}^B$

$$\tilde{\mathbf{E}}_D^{A(B)}(\mathbf{r} \in \Delta A) = \begin{cases} \tilde{\mathbf{E}}^{A(B)}(\mathbf{r}_\Sigma + 0^+\hat{\mathbf{n}}) & \text{for } h(\mathbf{r}_\Sigma) < 0 \\ \tilde{\mathbf{E}}^{A(B)}(\mathbf{r}_\Sigma - 0^+\hat{\mathbf{n}}) & \text{for } h(\mathbf{r}_\Sigma) > 0 \end{cases} \tag{13}$$

where $0^+$ is a small positive value. Note that $\tilde{\mathbf{E}}_D^{A(B)}(\mathbf{r}) = \tilde{\mathbf{E}}^{A(B)}(\mathbf{r})$ for $\mathbf{r} \notin \Delta A$.

Using Eq. (13), the coupling variation due to the nanohole deformation is given by

$$\Delta g = g_D - g = \iint_{\Delta A} \rho(r_x, r_y) \, dr_x dr_y, \tag{14}$$

where

$$\rho(r_x, r_y) = -\text{sgn}(h)(\varepsilon_s - \varepsilon_0) \int_0^t \tilde{\mathbf{E}}_D^A(\mathbf{r}) \cdot \tilde{\mathbf{E}}_D^B(\mathbf{r}) \, dr_z, \tag{15}$$

is the coupling variation density, which quantifies the impact of deformation on both coherent and dissipative couplings. In Eqs. (14-15), the integral is performed over the thickness $t$ of the photonic crystal membrane, along the cylinder defined by $\mathbf{r} = [r_x, r_y, r_z]$. $\varepsilon_s$ and $\varepsilon_0$ represent the permittivities of the semiconductor membrane material and air, while the sign function $\text{sgn}(h)$ is 1 for $h > 0$ and –1 otherwise.

Figure 4c visualizes the real and imaginary parts of $\rho(r_x, r_y)$. It serves as a lookup table. Each nanohole is represented by two concentric rings: the inner and outer rings correspond to inward and outward deformations, respectively. Generally, these deformations have opposite signs, as they induce opposite permittivity changes (see Eq. 15).

Using this figure, we now design a deformation to achieve zero coupling. The initial coupling value is $g_i = (2.36 - 0.089i)/Q_0$ with $Q_0 = 3500$ the quality factor of the individual cavity. This results in $Re(g_i) > 0$ and $Im(g_i) < 0$ (point A in Fig. 4d). From Fig. 4c, we see that decreasing the radii of holes labeled "2" significantly reduces $Re(g_i)$, while increasing the radii of holes labeled "5" increases $Im(g_i)$. A null coupling is achieved when $g_i + \Delta g_2 + \Delta g_5 = 0$, where $\Delta g_2 + \Delta g_5$ represents the total coupling variation due changes of hole raddii, $\Delta r_2$ and $\Delta r_5$. With two conditions on the real and imaginary parts of $g_i + \Delta g_2 + \Delta g_5$ and two degrees of freedom, $\Delta r_2$ and $\Delta r_5$, we obtain a unique solution: $\Delta r_2^{(pert)} = -0.155 r_0$ and $\Delta r_5^{(pert)} = 0.215 r_0$, where $r_0$ denotes the hole radius of the photonic crystal membrane.

Since first-order cavity perturbation theory is limited to small deformations, the coupling coefficient computed for the designed cavities is not exactly null. To refine the result, we further optimize $\Delta r_2$ and $\Delta r_5$ using a few iterations of gradient descent. With this refinement, we obtain $\Delta r_2 = -0.14 r_0$ and $\Delta r_5 = 0.19 r_0$. Figure 4d displays the evolution of the imaginary and real parts of $\Delta g_2$ and $\Delta g_5$ as the hole radii are gradually adjusted from $r_0$ to $0.86 r_0$ for hole "2" and from $r_0$ to $1.19 r_0$ for hole "5". Reference computational data obtained using MAN are marked with blue and red circles, while data generated using Eq. (14) and Fig. 4(c) appear as red and blue crosses. The small deviation between crosses and circles confirms the predictive accuracy of the perturbative cQNM approach with lookup tables.

To validate the zero-coupling condition through direct computations, we perform real-frequency simulations for three structures: the initial coupled cavities (labelled **A**), a structure (labelled **B**) where only the radii of holes '2' are modified, and the optimized zero-coupling structure (labelled **C**). For each case, we compute the electric field generated by a point electric dipole emitter placed at the center of the left cavity. The emitter operates at a real frequencies, $\omega = Re(\widetilde{\omega}_{ab})$, with $\widetilde{\omega}_{ab}$ the complex-frequency of the anti-bonding superQNM.

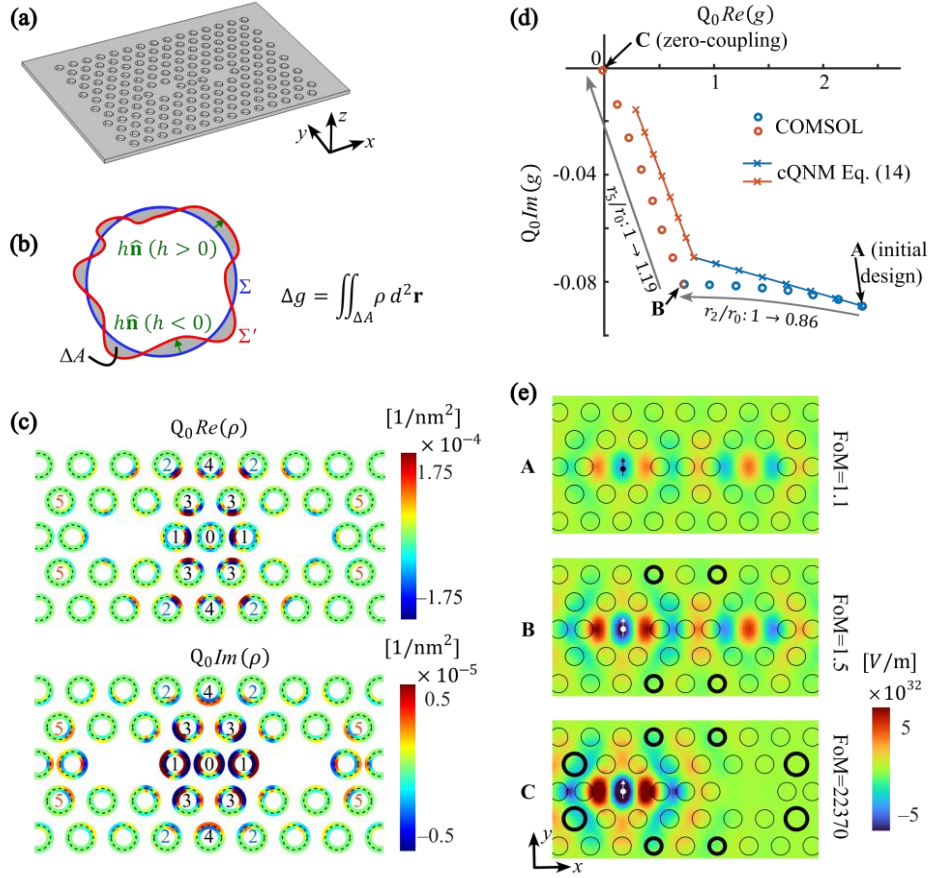

**Figure 4. Zero coupling between two very close photonic crystal cavities.** (a) Schematic of the coupled photonic-crystal cavities with lattice period $a = 420$ nm, membrane thickness $t = 252$ nm, permittivity $\varepsilon_s = 3.42$, and hole radius $r_0 = 0.29a$. The termination holes of each cavity are shifted by $0.2a$. (b) Illustration of a general nanohole deformation, where the boundary shifts from $\Sigma$ to $\Sigma'$, with inward ($h < 0$) and outward ($h > 0$) deformations. The influence of deformation on the coupling coefficient is determined by integrating over the deformed area $\Delta A$ in gray. (c) Real and imaginary components of the coupling variation density $\rho$, represented by two colorful rings corresponding to inward and outward deformations. (d) Evolution of the coupling coefficients as the radii of holes labeled '2' are reduced, and those of holes labeled '5' are increased. Circles represent results from COMSOL simulations, where $r_2$ varies from $r_0$ to $0.86 r_0$ and $r_5$ from $r_0$ to $1.19 r_0$. Crosses indicate predictions from Eq. (14), with $r_2$ ranging from $r_0$ to $0.845 r_0$ and $r_5$ from $r_0$ to $1.125 r_0$. (e) Maps of $Im(E_y)$ for three structures excited by a $y$-polarized electric-dipole emitter positioned at the center $\mathbf{r}_L$ of the left cavity. Structures **A**, **B**, and **C** correspond to hole radii $r_2 = r_5 = r_0$, $r_2 = 0.86 r_0$ & $r_5 = r_0$, and $r_2 = 0.86 r_0$ & $r_5 = 1.19 r_0$, respectively. The dipole emitter frequency is $Re(\widetilde{\omega}_{sb})$. The figure also shows the numerical values of the FoM. In (c) and (d), $Q_0 = 3500$, corresponding to the quality factor of the individual cavities.

Figure 4e presents the electric-field maps computed with COMSOL. In the initial coupled-cavity structure **A**, the field magnitudes inside both cavities are nearly identical, indicating strong coupling. In structure **B** where only the hole '2' radii are reduced, the field is more confined to the left cavity. Finally, in the zero-coupling structure **C**, the field in the right cavity is nearly zero, confirming the absence of coupling.

To quantify this decoupling with precision, we introduce a figure of merit defined as $\text{FoM} = \text{Im}(E_y(\mathbf{r}_L))/\text{Im}(E_y(\mathbf{r}_R))$, where $\text{Im}(E_y(\mathbf{r}_L))$ represents the local density of states at the center $\mathbf{r}_L$ of the left cavity and $\text{Im}(E_y(\mathbf{r}_R))$ represents the cross density of states at the center $\mathbf{r}_R$ of the right cavity. While this ratio is only 1 for the initial coupled cavities, it reaches an impressive 22,000 value for the optimized structure, therein demonstrating that the two cavities are effectively decoupled despite their close proximity.

The cQNM perturbation lookup table in Fig. 4(c), which reveals the influence of each hole on cavity coupling, is far from intuitive. Without the cQNM theory and this systematic perturbative approach, achieving zero coupling in such a compact structure would be extremely challenging.

### 4.3 Level attraction and repulsion

In coupled-resonator systems, the most common interaction is coherent coupling, which results in level repulsion or anticrossing. In more exotic scenarios, level attraction can also be observed [41-43,92]. These unusual phenomena arise in nonconservative systems that exchange energy through incoherent coupling.

We investigate this remarkable effect using a microwave Fabry-Perot stripe cavity coupled to a subwavelength Drude resonator via a stripe waveguide (Fig. 5a). This geometry was initially explored in [51], where the authors provide an interpretation based on CMT, employing a three-coupled-modes model, in which the dipolar mode of the Drude resonator and the Fabry-Perot stripe cavity are coupled through a dissipative mode of the stripe waveguide. Our goal is to demonstrate that the cQNM framework offers a clear and intuitive interpretation, without the need for fitting.

In line with [51], our COMSOL simulations (circles in Fig. 5b) of the superQNM eigenfrequencies reveal that both level repulsion and level attraction can be observed by adjusting the position (indicated by the orange dot in the insets) of the Drude resonator within the dielectric gap between the stripe waveguide and the metal backplane.

Rather than three modes and six fitting coupling coefficients in a 3×3 matrix, our model focuses on the direct coupling between 2 QNMs with a sign change in the coupling coefficient. It eliminates the need for fitting parameters and offers closed-form expressions by exploiting the fact that the Drude resonator is subwavelength, using exact expression for complex mode volume [53,55,66]. Moreover, the model can be readily extended to both classical and semi-classical descriptions of matter-vacuum hybrids in photonic cavities, where the hybrid ground state features strongly coupled photons [45,75].

We begin by computing the uncoupled QNMs using the **MAN** freeware, with a background permittivity defined by the stripe waveguide situated on top of the metal backplane, in the absence of stripe cavity labelled '$A$' and Drude resonator labelled '$B$'. Due to the proximity of the waveguide stripe, the cavity QNM field leaks significantly, leading to notable radiation loss ($\widetilde{\omega}^A = 2\pi \times (9.77 - 0.113i)$ GHz). The Drude resonator is modeled as a small ellipsoid with equal axis lengths of 0.61 mm along the $x$- and $y$-axes, and an axis length of 0.61 mm × ('Aspect Ratio') along the $z$-axis. The complex frequency of the electric dipolar mode is $\widetilde{\omega}^B = 2\pi \times (9.79 - 0.105i)$ GHz for an aspect ratio of 1. The QNM fields are shown in Fig. 5(c).

We then compute the coupling coefficients in the eigenproblem of Eq. (7). Given our choice for the permittivities $\varepsilon_\infty^A$ and $\varepsilon_\infty^B$, which are both equal to $\varepsilon_b$ at the cavity and Drude resonator locations (see the end of the caption of Fig. 5), the coupling coefficients $g^{AB}$ and $g^{BA}$ are null according to Eq. (6b). Consequently, the eigenfrequencies for the superQNMs are simply given by $\widetilde{\omega}_{+/-}^S = (\widetilde{\omega}^A + \widetilde{\omega}^B)/2 \pm \sqrt{\Delta^2/4 + \kappa^{AB}\kappa^{BA}}$, $\Delta = \widetilde{\omega}^A - \widetilde{\omega}^B$ being the frequency detuning. While the eigenfrequencies can be determined by numerically computing the coupling coefficients $\kappa^{AB}$ and $\kappa^{BA}$ as overlap integrals, see Eq. (5a), we aim to achieve *full analyticity* to highlight the strength of the approach.

To facilitate this, we make two key approximations. First, since $\widetilde{\omega}^A$ and $\widetilde{\omega}^B$ are relatively close in values, we assume that $\kappa^{AB} = \kappa^{BA} \equiv \kappa$. Second, given the small size of the Drude resonator '$B$', the

QNM field of the stripe cavity 'A' can be assumed to be uniform within the volume of the resonator, allowing us to introduce the concept of complex mode volume [45]. Using these approximations, Eq. (5a) leads to $\kappa = \kappa_0\, \mathbf{u} \cdot \tilde{\mathbf{E}}^A(\mathbf{r}_D)$, where $\kappa_0$ is an almost real quantity defined as $-\widetilde{\omega}^B \iiint_{\Omega_{res}^B} \varepsilon_L^B(\widetilde{\omega}_B)\tilde{\mathbf{E}}^B(\mathbf{r}) \cdot \mathbf{u}\, d^3\mathbf{r}$. The unit vector $\mathbf{u}$ is aligned with the polarization direction of $\tilde{\mathbf{E}}^A$ at the geometric center $\mathbf{r}_B$ of the Drude resonator. Specifically, $\mathbf{u} = \mathbf{e}_z$, the field is vertically polarized within the subwavelength-thick gap between the stripe waveguide and the backplane.

By introducing the complex mode volume of the stripe cavity $\tilde{V}_A^{-1} = 2\varepsilon_{gap}\tilde{\mathbf{E}}^A \cdot \tilde{\mathbf{E}}^A$, the eigenfrequencies of the superQNMs are expressed as

$$\widetilde{\omega}_{+/-}^S = (\widetilde{\omega}^A + \widetilde{\omega}^B)/2 \pm \sqrt{\Delta^2/4 + \kappa_0^2/(2\varepsilon_{gap}\tilde{V}_A)}, \tag{16}$$

where $\varepsilon_{gap}$ is the permittivity of the dielectric gap.

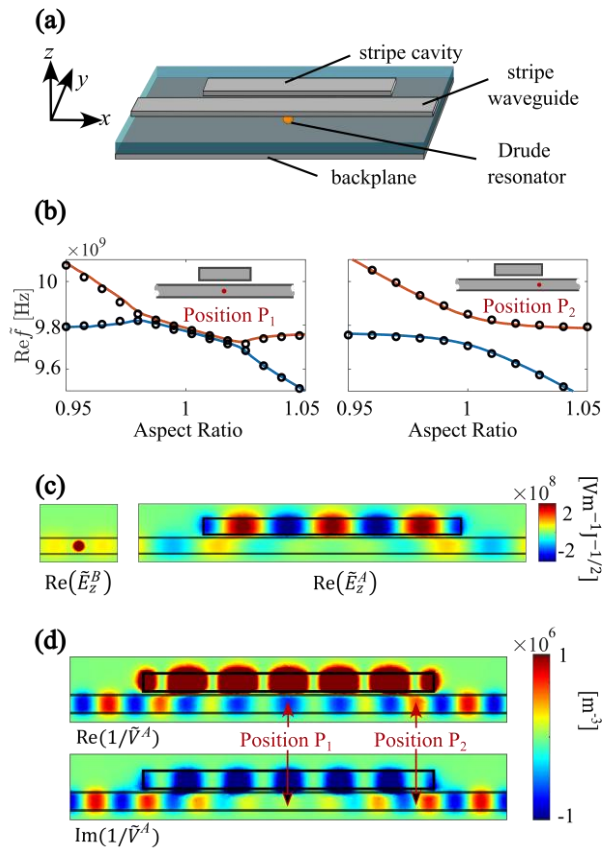

**Figure 5. Level attraction and repulsion between a Fabry-Perot stripe cavity and a tiny Drude resonator.** (**a**) Sketch of the coupled system. The cavity is labelled by '$A$' and the Drude resonator by '$B$'. (**b**) Real parts of the eigenfrequencies of the superQNMs when the Drude resonator (shown with an orange dot) is located close to the center of the cavity (position $P_1$) or close to its extremity (position $P_2$). Circles are COMSOL numerical results obtained by directly computing the QNMs of the coupled system with **MAN**. The solid curves are obtained with cQNM theory with Eq. (8). (**c**) Spatial maps of $\mathrm{Re}(\tilde{E}_z^A)$ and $\mathrm{Re}(\tilde{E}_z^B)$ (aspect ratio of 1) in the median plane of the dielectric gap. (**d**) Complex mode volume maps for the stripe cavity QNM in the median plane of the dielectric gap. Parameters are: cavity stripe length of 40 mm, stripe width and thickness of 2.5 mm and 0.5 mm, dielectric gap thickness of 1.5 mm and dielectric permittivity $\varepsilon_{gap} = 3.38\varepsilon_0$. The permittivities of the metal stripes and the Drude ellipsoid are approximated using a single-pole Drude model,

$\varepsilon_\infty^A = \varepsilon_0$, $\omega_{p,1} = 754$ GHz, $\gamma_1 = 0.0008\omega_{p,1}$ for the stripes and $\varepsilon_\infty^B = \varepsilon_{gap}$, $\omega_{p,1} = 754$ GHz, $\gamma_1 = 0.005\omega_{p,1}$ for the Drude resonator.

The solid blue and red curves in Fig. 5b are analytically calculated from Eq. (16) using the normalized QNM fields $\tilde{\mathbf{E}}^A$ and $\tilde{\mathbf{E}}^B$ computed with **MAN**, see Fig. 5c. These results align closely with the full-wave COMSOL computations (circles), confirming the predictive power of the cQNM theory and validating our hypotheses.

By observing the spatial distribution of the real and imaginary parts of $\tilde{V}_A$ in Fig. 5d, the nature of the coupling becomes immediately apparent. A positive mode volume (position $P_2$) corresponds to coherent coupling. For small detuning ($\Delta \approx 0$), it results in a level repulsion with distinct values for the superQNM frequencies, $\text{Re}(\tilde{\omega}_+^S) \neq \text{Re}(\tilde{\omega}_-^S)$. Reversely, a negative mode volume (position $P_1$) indicates purely dissipative coupling, leading to level attraction with $\text{Re}(\tilde{\omega}_+^S) \approx \text{Re}(\tilde{\omega}_-^S)$.

Lastly, although not the primary focus of this work, we note that beyond level attraction, negative mode volumes can give rise to several intriguing and counterintuitive effects. We refer interested readers to Suppl. Section S7, where we discuss three phenomena associated with the negative mode volume: an increase of quality factor with absorbing perturbers, a negative Purcell effect [55], and a driving source acting as a "sink" in the presence of a driving field.

## 5. Conclusion

In summary, we have developed a coupled-quasinormal-mode (cQNM) framework to analyze dissipative coupling between photonic and plasmonic resonators. Unlike the classical CMT approach, the framework offers closed-form expressions for the coupling coefficients and is not restricted to high-quality-factor modes in the weak-coupling limit. It may offer fully transparent interpretations of many phenomena of current interest in nanophotonics, e.g. zero-coupling between nearby cavities and coupling-induced level attraction.

The cQNM framework significantly advances our knowledge on dissipative coupling. It provides an effective method to predict the response of coupled systems and their supermodes using only the QNMs of the individual resonators, offering a microscopic insight into how the properties of individual resonators merge in complex systems. Furthermore, it involves simple algebra on small square matrices, enabling rapid parameter space exploration that is crucial for inverse photonic device design, particularly in LEGO-like scenarios involving prescribed resonators.

Prior works have already developed related frameworks with similar perspectives. Table 1 compares the present framework with previous ones. The evolution equation obtained in this work, see Eq. (2), is symbolically rewritten as $i|\dot{\boldsymbol{\beta}}\rangle = \begin{bmatrix} \tilde{\boldsymbol{\omega}} & \boldsymbol{\kappa} \\ \boldsymbol{\kappa} & \tilde{\boldsymbol{\omega}} \end{bmatrix} |\boldsymbol{\beta}\rangle - i\begin{bmatrix} 0 & \mathbf{g} \\ \mathbf{g} & 0 \end{bmatrix} |\dot{\boldsymbol{\beta}}\rangle + |\mathbf{F}\rangle$ to make the table more compact. As seen in column 3, our work differs from previous studies by deriving the coupling matrices $\boldsymbol{\kappa}$ and $\mathbf{g}$ in Eq. (2) without relying on approximations. Columns 4 and 5 highlight additional unique features of our framework, such as coupling through time derivatives of the modes ($\mathbf{g} \neq 0$) or the driving fields ($\mathbf{F} \propto \dot{\mathbf{E}}_D$), even for non-dispersive resonators.

Another distinctive feature of the present framework is a novel and highly accurate method for normalizing the superQNMs. This normalization enables the full application of the analytical toolbox provided by QNM theories to compute key physical quantities, such as the local density of states, mode volume, and perturbation-induced resonance shifts. The last column examines the feasibility of implementing accurate reconstructions, which we consider crucial as it demonstrates remarkable self-consistency and a high level of theoretical precision. Accurate reconstructions are only possible when formalism is free of significant approximations, avoids overly complex steps, and is fully developed.

The present framework also introduces several advantages. Notably, the supermode eigenvalue equation remains linear even in the presence of material dispersion—an important feature not emphasized in the referenced table and absent in one of the most advanced frameworks [62].

**Table 1. Overview of existing cQNM formalisms and comparison with the present formalism.**

| Reference | Dispersive materials | With or w/o Approx. | Time-derivative coupling | Time-derivative driving force | SuperQNM normalization | Reconstruction test |
|---|---|---|---|---|---|---|
| This work | yes | w/o | $\kappa \neq 0, g \neq 0$ | yes | yes | yes (Fig. 2) |
|  | no | w/o | $\kappa = 0, g \neq 0$ | yes |  |  |
| Vial, et al. [69] | no | w/o | $\kappa \neq 0, g \neq 0$ (3) | (3) | no | no (7) |
| Tasolamprou, et al. [74] | no | w/o | (4) | (3) | no | no |
| Cognée [70] | yes | w/o | $\kappa \neq 0, g \neq 0$ (3) | (3) | yes (5) | no (7) |
| Ren, et al. [72] | yes (1) | with | $\kappa \neq 0, g = 0$ (3) | (3) | no (6) | no |
| Tao, et al. [62,73] | yes | w/o | $\kappa \neq 0, g = 0$ (2) | (3) | no (6) | no |
| Muljarov [76] | yes | w/o | $\kappa \neq 0, g \neq 0$ (3) | (3) | no | no |
|  | no | w/o | $\kappa = 0, g \neq 0$ (3) | (3) | no | no (7) |

(1) However, only non-dispersive systems are considered.
(2) $\kappa$ is frequency-dependent for dispersive materials, making the super-QNM eigenproblem nonlinear.
(3) The present work is the only one that explicitly addresses time-dependent problems. We determine whether $\kappa$ and $g$ are null or non-null by analyzing the frequency-domain formula and comparing it to Eq. 5. The time dependence of the driving force term remains unspecified.
(4) The eigenvalues of the superQNM eigenequation are $(\widetilde{\omega}_m^S)^2$ instead of $\widetilde{\omega}_m^S$. We are not able to infer $\kappa$ and $g$.
(5) An approximate normalization of the superQNM is proposed, but the justification for this approximation is not provided.
(6) The normalization issue is circumvented with an analytical expression for the Green's function in the superQNM basis. In [72], the derived expression is approximate.
(7) The present work is the only one to provide a reconstruction of the optical response, e.g. the extinction cross section. Convergence tests for the superQNM eigenfrequencies were successfully reported in [69] and [76], and a similar test for the superQNM eigenvector in [70]. However, convergence was not achieved for the field inside the plasmonic resonators.

We stress that Table 1 is not intended to undervalue any previous work. Dissipative coupling between non-Hermitian electromagnetic bodies represents a real challenge for many decades and all studies are helpful towards achieving a comprehensive understanding. The Table is simply intended to highlight some peculiarities of the present framework that may require attention in future developments. We would also like to emphasize that some earlier studies did not include reconstruction tests, though this does not imply that their methods cannot achieve accurate reconstruction. For instance, we recently extended the approach in [69] to dispersive materials and successfully passed the reconstruction test outlined in Section 3, using the wire pair as a case study.

While this study focuses on linear, classical photonic systems, we foresee that the cQNM framework, much like the CMT, can be extended to systems involving nonlinear interactions, such as nonlinear couplings [36], optomechanical cooling [94], lasing with exceptional points [32], photonic switching [95], and strong coupling between quantum emitters and QNMs [96]. Moreover, while the present work focuses on coupling between systems composed of bulk materials, the approach can be extended to coupled systems involving 2D or 0D materials, such as van der Waals materials or quantum dots [97-99].

In the end, we would like to comment that the significant difference between the CMT Eq. (1) and the cQNM Eq. (2) raises questions about the validity of the CMT. The difference does not necessarily imply that CMT yields inaccurate results. In fact, the opposite is true, as the flexibility provided by fitting in CMT often allows for accurate modeling. A thorough investigation of the restrictions under which Eqs. (1) and (2) produce comparable results is beyond the scope of this work. However, such investigation has been recently done in the context of the evolution equation of resonators, focusing on the time-derivative driving term. It has been found that CMT remains accurate when the envelopes of the driving pulses are slowly varying [67] and when the QNM frequency matches the central frequency of the pulse [68]. We anticipate that similar conclusions apply to the present study of coupled resonators for essentially the same reasons.

## Acknowledgements

This work began over a decade ago. The authors extend their gratitude to Mathias Perrin, Kévin Cognée, Wei Yan, and Benjamin Rousseaux for their contributions at various stages. They also thank Femius Koenderink, Thomas Christopoulos, and Alejandro Giacomotti for insightful discussions. PL acknowledges financial support from the WHEEL project (ANR-22CE24-0012-03).

# Supplementary Materials: Dissipative Coupling in Photonic and Plasmonic Resonators

Tong Wu,[1] and Philippe Lalanne[1]*

[1]Laboratoire Photonique, Numérique et Nanosciences (LP2N), IOGS-University of Bordeaux-CNRS, 33400 Talence cedex, France

*philippe.lalanne@institutoptique.fr

## S1. Extension of the theory to the case of multiple resonators

Equations (5), (7), and (8) in the main text can be naturally extended to systems composed of more than two resonators. Labeling the resonators as $A, B, C, \ldots$ we consider each resonator to be driven by both the incident wave and the scattered fields of all other resonators. In this case, the coupled quasi-normal mode (cQNM) equation generalizes to

$$\widehat{\mathbf{H}}|\boldsymbol{\alpha}(\omega)\rangle = \omega\widehat{\mathbf{D}}|\boldsymbol{\alpha}(\omega)\rangle + |\mathbf{F}(\omega)\rangle, \tag{S1.1}$$

which retains the same form as Eq. (5) but with redefined matrices to accommodate multiple resonators. $\widehat{\mathbf{H}}, \widehat{\mathbf{D}}$ are given by

$$\widehat{\mathbf{H}} = \begin{bmatrix} \widetilde{\boldsymbol{\omega}}^A & \boldsymbol{\kappa}^{AB} & \boldsymbol{\kappa}^{AC} & \ldots \\ \boldsymbol{\kappa}^{BA} & \widetilde{\boldsymbol{\omega}}^B & \boldsymbol{\kappa}^{BC} & \ldots \\ \boldsymbol{\kappa}^{CA} & \boldsymbol{\kappa}^{CB} & \widetilde{\boldsymbol{\omega}}^C & \vdots \\ \vdots & \vdots & \ldots & \ddots \end{bmatrix}, \tag{S1.2}$$

and

$$\widehat{\mathbf{D}} = \begin{bmatrix} \mathbf{I} & \mathbf{g}^{AB} & \mathbf{g}^{AC} & \ldots \\ \mathbf{g}^{BA} & \mathbf{I} & \mathbf{g}^{BC} & \ldots \\ \mathbf{g}^{CA} & \mathbf{g}^{CB} & \mathbf{I} & \vdots \\ \vdots & \vdots & \ldots & \ddots \end{bmatrix}, \tag{S1.3}$$

where $\boldsymbol{\kappa}^{XY}$ and $\mathbf{g}^{XY}$ (with $X, Y = A, B, C \ldots$) are the coupling matrices. Their elements are defined analogously to the two-resonator case in Eq. (6) of the main text.

The driven term is defined as $|\mathbf{F}(\omega)\rangle = \left[\alpha_{1,D}^A(\widetilde{\omega}_1^A - \omega), \alpha_{2,D}^A(\widetilde{\omega}_2^A - \omega) \ldots \alpha_{1,D}^B(\widetilde{\omega}_1^B - \omega), \alpha_{2,D}^B(\widetilde{\omega}_2^B - \omega) \ldots \alpha_{1,D}^C(\widetilde{\omega}_1^C - \omega), \alpha_{2,D}^C(\widetilde{\omega}_2^C - \omega) \ldots\right]^T$, with $\alpha_{m,D}^X$ ($X$ taking $A, B, C, \ldots$) computed by Eq. (4) in the main text with $\mathbf{E}_s^X \equiv \mathbf{0}$.

The vector containing all modal excitation coefficients is $|\boldsymbol{\alpha}\rangle = \left[\alpha_1^A, \alpha_2^A \ldots \alpha_1^B, \alpha_2^B \ldots \alpha_1^C, \alpha_2^C \ldots\right]^T$. This vector can be determined either through direct matrix inversion $|\boldsymbol{\alpha}(\omega)\rangle = \left[\widehat{\mathbf{H}} - \omega\widehat{\mathbf{D}}\right]^{-1}|\mathbf{F}(\omega)\rangle$ or via the eigenstate expansion, $|\boldsymbol{\alpha}\rangle = \sum_m c_m^S(\omega)|\boldsymbol{\alpha}_m^S\rangle$. As in the two-resonator case, the eigenvectors $|\boldsymbol{\alpha}_m^S\rangle$ are obtained by solving the eigenvalue problem

$$\widehat{\mathbf{H}}|\boldsymbol{\alpha}_m^S\rangle = \widetilde{\omega}_m^S \widehat{\mathbf{D}}|\boldsymbol{\alpha}_m^S\rangle, \tag{S1.4}$$

and the supermode excitation coefficient $c_m^S(\omega)$ is given by

$$c_m^S(\omega) = \frac{1}{(\widetilde{\omega}_m^S - \omega)} \frac{\langle \boldsymbol{\alpha}_m^S|\mathbf{F}\rangle}{\langle \boldsymbol{\alpha}_m^S|\widehat{\mathbf{D}}|\boldsymbol{\alpha}_m^S\rangle}. \tag{S1.5}$$

Once $|\boldsymbol{\alpha}\rangle$ is known, the total field is computed as the sum of the scattered fields from all resonators and the incident field: $\mathbf{E}_{tot}(\mathbf{r}, \omega) = \mathbf{E}_D(\mathbf{r}, \omega) + \mathbf{E}_s^A(\mathbf{r}, \omega) + \mathbf{E}_s^B(\mathbf{r}, \omega) + \mathbf{E}_s^C(\mathbf{r}, \omega) + \cdots$, where $\mathbf{E}_s^X(\mathbf{r}, \omega) = \sum_m \alpha_m^X(\omega) \widetilde{\mathbf{E}}_m^X(\mathbf{r})$, ($X$ representing $A, B, C, \ldots$).

## S2. Normalization of the superQNMs

In this section, we study the normalization of the superQNM. As a side note, we emphasize that the theory developed in Sections 2 and S1, intrinsically allows the computation of the optical responses of the coupled system using superQNMs, and therefore normalization is not necessarily required.

However, QNM normalization holds intrinsic physical significance [1-3]: the field intensity of the normalized QNMs tells the QNM capability to respond to a driving field [4-5]; the phase encodes the non-Hermitian character of QNMs and determines the line shape of the Purcell factor [1,6]; the inverse of the square of the normalized QNM corresponds to the complex mode volume [7] ($\tilde{V}^{-1} \propto \tilde{\mathbf{E}}^2$). Additionally, using normalized QNMs enables the direct application of formulas derived in previous works to compute physical quantities [3,8], such as third- or second-harmonic generations, the local density of states (LDOS), or perturbation induced resonance shift.

QNM normalization has been debated during almost 10 years [9] and the debate has been clarified in [9] and then [2]. Two rigorous strategies can be adopted to obtain the normalization factor: the PML normalization approach [1-3,10] and the pole-response approach [2,3,5].

Before discussing the normalization, let us first introduce the superQNM field expressed in terms of the QNM fields of the uncoupled structures

$$\tilde{\mathbf{E}}_m^S(\mathbf{r}) = \sum_n \alpha_{m,n}^A \tilde{\mathbf{E}}_n^A(\mathbf{r}) + \alpha_{m,n}^B \tilde{\mathbf{E}}_n^B(\mathbf{r}),  \qquad (S2.1)$$

where $\alpha_{m,n}^A$ and $\alpha_{m,n}^B$ are elements of the eigenvectors $|\alpha_m^S\rangle = [\alpha_{m,1}^A, \alpha_{m,2}^A \ldots \alpha_{m,1}^B, \alpha_{m,2}^B \ldots]^T$ with eigenfrequency $\tilde{\omega}_m^S$ obtained by solving Eq. (7).

Since $\tilde{\mathbf{E}}_m^S$ is not yet normalized, we define the normalized field as

$$\tilde{\mathbf{E}}_m^{SN}(\mathbf{r}) = \pm \tilde{\mathbf{E}}_m^S(\mathbf{r})/\sqrt{QN_m}, \qquad (S2.2)$$

where the plus and minus sign is unimportant since any physical quantity of interest, such as the modal contribution to a QNM expansion or the mode volume, is expected to involve $\left(\tilde{\mathbf{E}}_m^{SN}\right)^2$.

In the following subsections, we explain how to compute the normalization factor $QN_m$ using both the PML normalization and pole-response approaches.

## S2.1 PML normalization

In the PML normalization approach [1-4], the normalization factor is computed by integrating both the physical and PML domains (refer to Fig. 1 for the notations of $\Omega_{phy}$ and $\Omega_{PML}$):

$$QN_m = \iiint_{\Omega_{phy} \cup \Omega_{PML}} \tilde{\mathbf{E}}_m^S \cdot \frac{\partial(\omega\varepsilon)}{\partial\omega} \tilde{\mathbf{E}}_m^S - \tilde{\mathbf{H}}_m^S \cdot \frac{\partial(\omega\mu)}{\partial\omega} \tilde{\mathbf{H}}_m^S \mathrm{d}^3\mathbf{r}. \qquad (S2.3)$$

The superQNMs $\tilde{\mathbf{E}}_m^S$ and $\tilde{\mathbf{H}}_m^S$ are computed with the knowledge of the eigenvectors $|\alpha_m^S\rangle$ and the QNMs of the isolated resonators.

## S2.2 Pole response normalization

Alternatively, the pole-response approach [2,3,5] can be employed. This method exploits the fact that, for complex frequencies $\omega$ sufficiently close to the complex QNM eigenfrequency $\tilde{\omega}_m^S$, the field $\mathbf{E}_s(\mathbf{r}, \omega)$ scattered by the resonator is proportional to the normalized QNM field $\tilde{\mathbf{E}}_m^{SN}$ with a known excitation coefficient $\alpha_m^S$:

$$\lim_{\omega \to \tilde{\omega}_m^S}(\omega - \tilde{\omega}_m^S)\mathbf{E}_s(\mathbf{r}, \omega) = \lim_{\omega \to \tilde{\omega}_m^S}(\omega - \tilde{\omega}_m^S)\alpha_m^S(\omega)\tilde{\mathbf{E}}_m^{SN}(\mathbf{r}), \qquad (S2.4)$$

where

$$\lim_{\omega \to \tilde{\omega}_m^S} \alpha_m^S(\omega) \equiv \tilde{\omega}_m^S(\tilde{\omega}_m^S - \omega)^{-1} \iiint_{\Omega_r^A \cup \Omega_r^B} \Delta\varepsilon(\tilde{\omega}_m^S)\tilde{\mathbf{E}}_m^{SN}(\mathbf{r}) \cdot \mathbf{E}_D(\mathbf{r}, \tilde{\omega}_m^S) \, \mathrm{d}^3\mathbf{r}. \qquad (S2.5)$$

Here $\Delta\varepsilon = \Delta\varepsilon^X$ for $\mathbf{r} \in \Omega_r^X$ (with $X \in \{A, B\}$). By substituting $\mathbf{E}_s$ in Eq. (S2.4) with $\mathbf{E}_s(\mathbf{r}, \omega) = \sum_m \alpha_m^A(\omega)\tilde{\mathbf{E}}_m^A(\mathbf{r}) + \alpha_m^B(\omega)\tilde{\mathbf{E}}_m^B(\mathbf{r})$ obtained from the cQNM theory, and $\tilde{\mathbf{E}}_m^{SN}$ with Eq. (S2.2), the normalization factor $QN_m$ can be determined.

If the uncoupled QNMs form a complete set, the value of $QN_m$ should be independent of the driving field $\mathbf{E}_D$. In the practical implementation of the method, $\mathbf{E}_D$ can be chosen as any function that guarantees the integration in Eq. (S2.5) does not vanish.

### S2.3 Pole response normalization with approximations

The two previous approaches are rigorous; however, they involve multiple steps and lack analytical simplicity. To preserve full analyticity in the cQNM treatment, it is convenient to approximate $QN_m$ as

$$QN_m \approx \langle \boldsymbol{\alpha}_m^S | \hat{\mathbf{D}} | \boldsymbol{\alpha}_m^S \rangle, \tag{S2.6}$$

provided the following two conditions hold

1. The superQNM eigenfrequencies $\tilde{\omega}_m^S$ remain close to those of the uncoupled resonators, i.e., $\tilde{\omega}_m^S \approx \tilde{\omega}_n^A \approx \tilde{\omega}_n^B$.
2. The permittivities of the two resonators are approximately equal.

For most systems where only a few spectrally close modes dominate—such as those in Figs. 3–5—Eq. (S2.6) provides a good approximation. However, this approach becomes unreliable when a large number of modes spanning a broad frequency range are included in the cQNM equation, as seen in Fig. 2. In such cases, rigorous normalization is required.

A formal proof of Eq. (S2.6) is presented in Section S3.

## S3. Approximate normalization to preserve full analyticity in cQNM method

To demonstrate Eq. (S2.6), we adopt the Pole Response Normalization Approach. In this approach, the driven field $\mathbf{E}_D$ in Eq. (S2.5) is arbitrary. We consider that $\mathbf{E}_D$ is the field emitted by an electric dipole current located at $\mathbf{r}_p$ with dipolar moment $\mathbf{p}$

$$\mathbf{E}_D(\mathbf{r}, \omega) = -i\omega \overline{\mathbf{G}}(\mathbf{r}, \mathbf{r}_p, \omega) \cdot \mathbf{p}, \tag{S3.1}$$

where $\overline{\mathbf{G}}$ is Green's tensor for the background medium in the absence of the two resonators.

When the dipole current oscillates with a frequency $\omega \approx \tilde{\omega}_m^S$, Eqs. (S2.4) and (S2.5) lead to

$$\lim_{\omega \to \tilde{\omega}_m^S}(\omega - \tilde{\omega}_m^S)\mathbf{E}_s(\mathbf{r}, \omega) = -\tilde{\mathbf{E}}_m^{SN}(\mathbf{r})\tilde{\omega}_m^S \sum_{X=A,B} \iiint_{\Omega_r^X} \Delta\varepsilon^X(\tilde{\omega}_m^S)\tilde{\mathbf{E}}_m^{SN}(\mathbf{r}') \cdot \mathbf{E}_D(\mathbf{r}', \tilde{\omega}_m^S) \, d^3\mathbf{r}' \tag{S3.2}$$

Substituting Eq. (S2.2) into the right-hand side (R.H.S) of Eq. (S3.2) yields

$$\text{R.H.S} = -QN_m^{-1}\tilde{\mathbf{E}}_m^S(\mathbf{r})\tilde{\omega}_m^S \sum_{X=A,B} \iiint_{\Omega_r^X} \Delta\varepsilon^X(\tilde{\omega}_m^S)\tilde{\mathbf{E}}_m^S(\mathbf{r}') \cdot \mathbf{E}_D(\mathbf{r}', \tilde{\omega}_m^S) \, d^3\mathbf{r}'. \tag{S3.3}$$

Using Eq. (8), the left-hand side (L.H.S) of Eq. (S3.2) can be rewritten as

$$\text{L.H.S} = \lim_{\omega \to \tilde{\omega}_m^S} -\tilde{\mathbf{E}}_m^S(\mathbf{r})\langle \boldsymbol{\alpha}_m^S | \mathbf{F}(\omega) \rangle \langle \boldsymbol{\alpha}_m^S | \hat{\mathbf{D}} | \boldsymbol{\alpha}_m^S \rangle^{-1}. \tag{S3.4}$$

At this stage, no approximations have been applied. The following derivation introduces approximations to prove Eq. (S2.6). Readers interested only in the final result can note that Eq. (S2.6) holds under the following three approximations, which are valid when the conditions outlined in Section S2.3 are satisfied:

   **Approx. 1.** $\tilde{\omega}_n^A \approx \tilde{\omega}_m^S$ and $\tilde{\omega}_n^B \approx \tilde{\omega}_m^S$.

   **Approx. 2.** Because of Approx. 1, the Green tensors at different frequencies are approximately equal, $\overline{\mathbf{G}}(\mathbf{r}, \mathbf{r}_p, \tilde{\omega}_m^S) \approx \overline{\mathbf{G}}(\mathbf{r}, \mathbf{r}_p, \tilde{\omega}_n^A) \approx \overline{\mathbf{G}}(\mathbf{r}, \mathbf{r}_p, \tilde{\omega}_n^B)$.

**Approx. 3.** The left and right eigenvectors are colinear, $\langle\widetilde{\Psi}_m^S|^T = |\widetilde{\Psi}_m^S\rangle$. This follows from the coupling matrices $\widehat{H}$ and $\widehat{D}$ can be approximately viewed to be symmetric under the consitions given by section S2.3.

Using Eq. (6c), the term $\langle\boldsymbol{\alpha}_m^S|\mathbf{F}(\omega)\rangle$ can be expressed as

$\lim_{\omega\to\widetilde{\omega}_m^S}\langle\boldsymbol{\alpha}_m^S|\mathbf{F}(\omega)\rangle$
$\approx^a \sum_{X=A,B;n=1,2,\ldots} \tilde{\alpha}_{m,n}^{X,L} \iiint_{\Omega_r^X} \widetilde{\mathbf{E}}_n^X(\mathbf{r}) \cdot [\widetilde{\omega}_n^X \Delta\varepsilon(\widetilde{\omega}_n^X)] \mathbf{E}_D(\mathbf{r},\widetilde{\omega}_m^S) d^3\mathbf{r}$
$\approx^b \sum_{X=A,B;n=1,2,\ldots} -i\widetilde{\omega}_n^X \tilde{\alpha}_{m,n}^{X,L} \iiint_{\Omega_r^X} \widetilde{\mathbf{E}}_n^X(\mathbf{r}) \cdot [\widetilde{\omega}_n^X \Delta\varepsilon(\widetilde{\omega}_n^X)] \overline{\mathbf{G}}(\mathbf{r},\mathbf{r}_p,\widetilde{\omega}_n^X) \cdot \mathbf{p} d^3\mathbf{r}$
$=^c \sum_{X=A,B;n=1,2,\ldots} -i\widetilde{\omega}_n^X \tilde{\alpha}_{m,n}^{X,L} \iiint_{\Omega_r^X} \mathbf{p} \cdot \overline{\mathbf{G}}(\mathbf{r}_p,\mathbf{r},\widetilde{\omega}_n^X) \cdot [\widetilde{\omega}_n^X \Delta\varepsilon(\widetilde{\omega}_n^X)] \widetilde{\mathbf{E}}_n^X(\mathbf{r}) d^3\mathbf{r}$
$=^d \sum_{X=A,B;n=1,2,\ldots} \widetilde{\omega}_n^X \tilde{\alpha}_{m,n}^{X,L} \mathbf{p} \cdot \widetilde{\mathbf{E}}_n^X(\mathbf{r}_p)$
$\approx^e \widetilde{\omega}_m^S \mathbf{p} \cdot \widetilde{\mathbf{E}}_m^S(\mathbf{r}_p),$ (S3.5)

where $\tilde{\alpha}_{m,n}^{X,L}$ is the element of the left eigenvector $\langle\boldsymbol{\alpha}_m^S|$, defined such that $\langle\boldsymbol{\alpha}_m^S| = [\tilde{\alpha}_{m,1}^{A,L}, \tilde{\alpha}_{m,2}^{A,L}, \ldots, \tilde{\alpha}_{m,1}^{B,L}, \tilde{\alpha}_{m,2}^{B,L}, \ldots]$. The technical details of the steps are given as follows, where we use '●' and '○' to denote steps with and without approximations, respectively.

- Step a: considering **Approx. 1**, ignore the term proportional to $(\widetilde{\omega}_n^X - \widetilde{\omega}_m^S)$.
- Step b: use the definition Eq. (S3.1) and the **Approx. 1** and **2**.
- Step e: use **Approx. 1** and **3**, and Eq. (S2.1).
- Step c: use the reciprocity of Green's tensor $\overline{\mathbf{G}}(\mathbf{r},\mathbf{r}_p,\omega) = \overline{\mathbf{G}}^T(\mathbf{r}_p,\mathbf{r},\omega)$.
- Step d: use the Born theorem $\widetilde{\mathbf{E}}_n^X(\mathbf{r}) = -i\widetilde{\omega}_n^X \iiint_{\Omega_r^X} \overline{\mathbf{G}}(\mathbf{r},\mathbf{r}',\widetilde{\omega}_n^X) \cdot \Delta\varepsilon^X(\widetilde{\omega}_n^X) \widetilde{\mathbf{E}}_n^X(\mathbf{r}') d^3\mathbf{r}'$.

For the R.H.S.,

R.H.S.
$=^f iQN_m^{-1} \widetilde{\mathbf{E}}_m^S(\mathbf{r}) \widetilde{\omega}_m^S \sum_{X=A,B} \iiint_{\Omega_r^X} \widetilde{\omega}_m^S \Delta\varepsilon^X(\widetilde{\omega}_m^S) \widetilde{\mathbf{E}}_m^S(\mathbf{r}') \cdot \overline{\mathbf{G}}(\mathbf{r}_p,\mathbf{r}',\widetilde{\omega}_m^S) \cdot \mathbf{p} \, d^3\mathbf{r}'$
$=^g -\widetilde{\omega}_m^S Q N_m^{-1} \widetilde{\mathbf{E}}_m^S(\mathbf{r})[\mathbf{p}\cdot \widetilde{\mathbf{E}}_m^S(\mathbf{r}_p)].$ (S3.6)

- Step f: substitute $\mathbf{E}_D(\mathbf{r}',\widetilde{\omega}_m^S)$ with Eq. (S3.1)
- Step g: adopt the Born theorem.

By equating the L.H.S. and R.H.S., the normalization factor is shown to be equal with the inner product, Eq. (S2.6). Notably, the normalization expression has been previously used in [11], but no formal justification was provided there.

## S4. Supplementary results for the numerical convergence test in Fig.2b

In Fig. 2b of the main text, we presented the coupling coefficients between five dominant QNMs of the uncoupled nanowires. Here, we provide the corresponding field distributions and eigenfrequencies of these modes.

The results are displayed in Fig. S1. Due to the system's symmetry, the QNM fields of the two resonators exhibit mirror symmetry and share identical eigenfrequencies. Four out of the five modes are QNMs with electric fields that are well confined within the nanowire. In contrast, the mode with an eigenfrequency of 433.7–294.4i THz is identified as a PML mode, characterized by an electric field primarily distributed outside the resonator.

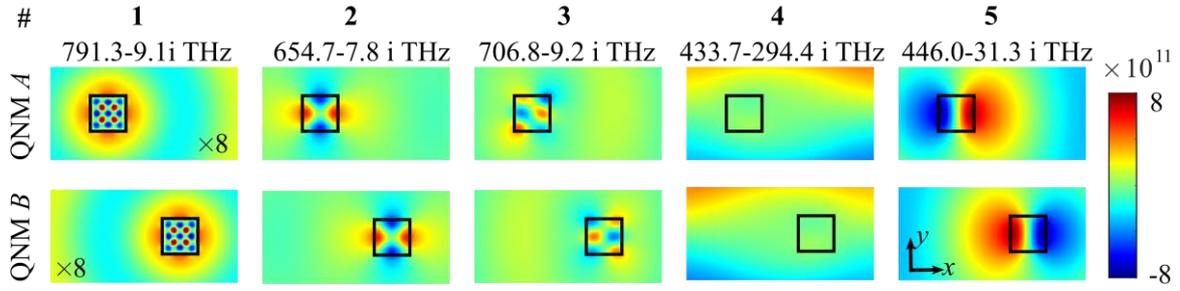

**Figure S1.** Eigenfrequencies ($\widetilde{\omega}^{A(B)}/2\pi$) and normalized electric field z-component ($\widetilde{E}_z^{A(B)}$) in xy-plane for the five dominant QNMs of the single nanowire shown in Fig. 2b. The normalized field of the first QNM is significantly weaker than the others and is therefore multiplied by a factor of 8 for visual clarity. The modes displayed in columns 1, 2, 3, and 5 correspond to QNMs, while the mode in column 4 is identified as a PML mode.

## S5. Ultrafast cQNM solvers

In this section, we demonstrate how the cQNM framework can predict the optical responses of 'lego' structures made of interacting resonators using straightforward algebra involving small-dimensional matrices. Once the QNMs of individual resonators are calculated using a solver like **MAN** [12], the superQNMs and optical responses of various 'lego' configurations can be evaluated in seconds. This significantly reduces the computational cost of exploring large parameter spaces, making the design of dimers, trimers, and polymers much more efficient.

To illustrate our approach, we consider plasmonic molecules composed of nanorods with distinct shapes (Fig. S2). Unlike the dimer in Fig. 2, where PML modes are necessary, the dominant near-field coupling allows for reasonably accurate results without including PML modes. For simplicity and efficiency, we focus on a single dipolar QNM per nanorod, demonstrating the resource-saving potential of the method. Greater precision, comparable to that in Fig. 2, can be achieved by incorporating additional QNMs.

We test three geometries: two dimers with orthogonal or parallel rods (Figs. S2(a-b)) and a trimer with one short rod and two long rods (Fig. S2(c)). All geometries maintain a gap width of 20 nm between rods. For the trimer, we employ cQNM formulas for a three-body system, derived directly from the two-body system equations (Section S1).

Using the eigenfrequencies and QNM fields of the uncoupled rods, we construct 2×2 (for the dimers) or 3×3 (for the trimer) coupling matrices. The coupling coefficients $\kappa^{XY}$ are presented in the first column of Fig. S2. Generally, the real part of the coupling coefficient is significantly larger than the imaginary part, indicating primarily coherent coupling. Notable exceptions are observed when the rods are parallel, as evident in the complex coefficients $\kappa^{A1A2}$ and $\kappa^{A2A1}$ in Fig. S2c.

Diagonalizing the homogeneous Eq. (7) provides the superQNM eigenfrequencies, displayed in the second column of Fig. S2. The width of the horizontal bars corresponds to the imaginary part of the eigenfrequency.

Finally, the total extinction cross-section is calculated, summing contributions from all superQNMs under plane wave illumination (polarization and incidence directions shown in insets of Figs. S2(a-c). The summed results (solid blue curves) align closely with reference data computed directly using COMSOL (black circles).

The examples above demonstrate that the cQNM method is remarkably robust and effective for arranging particles and controlling their interactions, provided the size and shape of the particles remain unchanged. Since hybrid responses are generally governed by a few dominant QNMs, solving the direct problem involves diagonalizing an ultrasmall-dimensional matrix (Eq. (5)) with analytically known coefficients. This simplifies the traditionally complex task of computing gradients in large parameter spaces, which often requires finite-difference schemes or adjoint methods, making it significantly more manageable.

Furthermore, advanced high-order QNM perturbation frameworks exist for calculating new QNMs of significantly deformed particles by leveraging the initial QNMs of unperturbed particles [13,14]. Integrating these perturbation frameworks with the current method would represent a significant advancement in optimizing the design of 'photonic or plasmonic molecules.'

In closing, we highlight a technical issue related to the reconstruction of the extinction in the systems shown in Fig. S2. The extinction can be computed using the formula: $\sigma_{ext} = \frac{1}{2S_0} \sum_{X=A,B} \iiint_{\Omega_{res}^X} \text{Im}[\omega \Delta \varepsilon^X \mathbf{E}_{tot} \cdot \mathbf{E}_D^*] d^3\mathbf{r}$ with $S_0 = 1/2\sqrt{\varepsilon_0/\mu_0}\,|\mathbf{E}_{inc}|^2$ the energy flow of the incident plane wave. Note that $\mathbf{E}_D$ is obtained by exciting the hosting medium with $\mathbf{E}_{inc}$, and the two fields are not equal when the hosting medium is non-uniform.

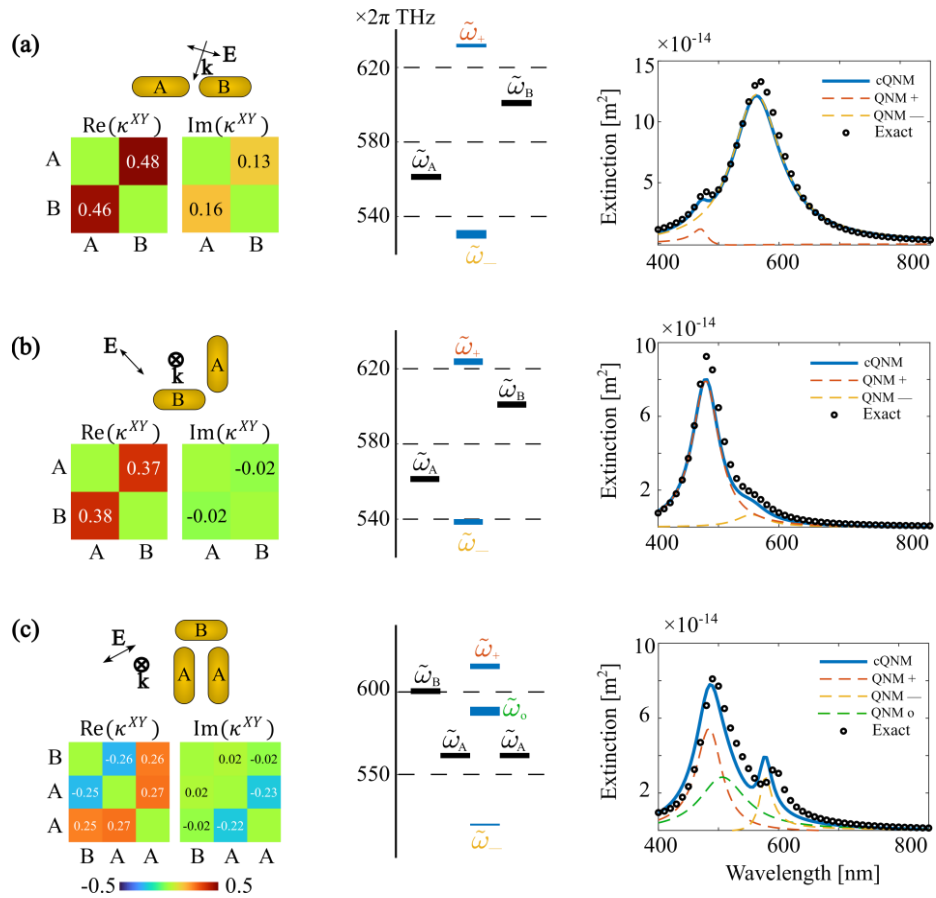

**Figure S2. Fast computation of the superQNM eigenfrequencies and optical responses for nanorod molecules using the cQNM method**. The system, embedded in an air background, consists of two types of nanorods with lengths $l_A = 110$ nm and $l_B = 100$ nm, each with a diameter of 40 nm. Three configurations are examined. (**a**)-(**b**) Dimers with inter-rod angles of 0° and 90°, and (**c**) dolmen-like structure with two identical rods. In all configurations, the gap between rods is 20 nm. The polarization and direction of the incident plane wave are shown in the insets. The incident angle in (**a**) is $80°$ for (**a**), and the electric fields are polarized in $45°$ and $30°$ degrees for (**b**) and (**c**). (**a**)-(**c**) The first column displays the coupling coefficients $\kappa_{mn}^{XY}$. The second column provides an energy level diagram, illustrating mode hybridization: black bars represent the eigenfrequencies of the uncoupled dipolar QNMs; blue bars show the hybrid modes, with thickness indicating the imaginary part of the eigenfrequency. The third column presents the extinction spectra, and the dashed lines indicate the contributions of every superQNMs, obtained through projection. The summed contributions closely match the reference COMSOL real-frequency

simulation data (circles). Detailed models and codes for reproducing this figure are available in **MAN** version V9 and following ones [12].

According to Eq. (3), the scattered fields of two resonators, $\mathbf{E}_s^A$ and $\mathbf{E}_s^B$, can be computed using the QNMs of the individual resonators. The total field is then given by: $\mathbf{E}_{tot} = \mathbf{E}_s^A + \mathbf{E}_s^B + \mathbf{E}_D$.

In principle, this computation is exact if a complete set of QNMs of the individual resonators is considered. However, as reported in [12], discontinuities in the perpendicular component of the electric field across the resonator boundary can introduce inaccuracies when using Eq. (3)—unless a large set of static modes is included.

This issue can be mitigated using the auxiliary field reconstruction method, which yields

$$\mathbf{E}_{tot}(\omega, \mathbf{r}) = \sum_m \frac{\widetilde{\omega}_m^X \varepsilon_L^X(\widetilde{\omega}_m^A)}{\omega \varepsilon_L^X(\omega)} \alpha_m^X(\omega) \widetilde{\mathbf{E}}_m^X(\mathbf{r}) \text{ for } \mathbf{r} \in \Omega_{res}^X. \tag{S5}$$

Unlike Eq. (3), this approach directly computes the total field as a weighted sum of the QNMs of the individual resonator to which **r** belongs, remaining independent of the other resonator and the incident field. Our previous studies during the past few years [4,15,16] have shown that when only a few QNMs are considered, Eq. (S5) provides significantly improved accuracy.

The results in Fig. S2 are obtained with Eq. (S5). We found that if Eq. (3) were used instead, the accuracy would be significantly worse.

## S6. QNMs and structures of the uncoupled cavities

Figure S3 shows the QNM field distributions and structural configurations of the uncoupled photonic cavities in Section 4.2. Cavities *A* and *B* are defined as photonic crystal structures without the shifted and missing holes on the right and left sides, respectively.

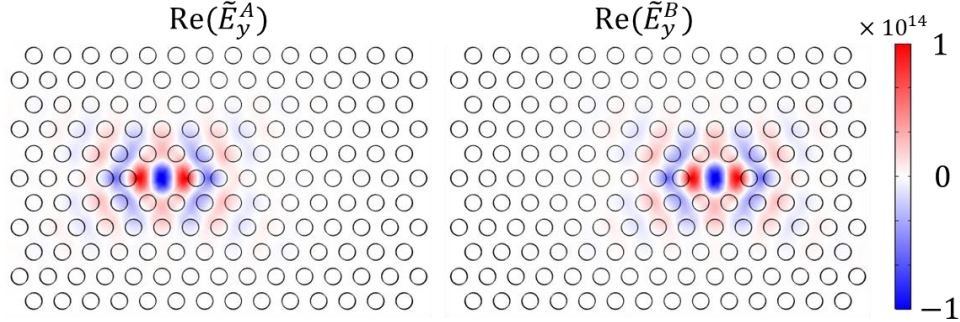

**Figure S3.** QNM field distributions for cavities *A* and *B* discussed in Section 4.2.

## S7. Intriguing effects caused by the negative mode volumes

Although not the primary focus of this work, it is worth emphasizing that negative mode volumes can give rise to counterintuitive effects beyond level attraction. First, according to first-order QNM perturbation theory, the frequency shift $\Delta\widetilde{\omega}$ due to a permittivity change $\Delta\varepsilon$ induced by a subwavelength perturber is $\Delta\widetilde{\omega} = -\widetilde{\omega}\Delta\varepsilon/(2\widetilde{V})$ [7], which suggests that the quality factor can actually increase when lossy material, $Im(\Delta\varepsilon) > 0$, is added to regions with negative mode volumes. This is because the absorption loss caused by the perturber reduces radiation loss. A similar phenomenon has been observed in active systems with gain materials, where adding loss materials can lower the lasing threshold [17], though without a detailed quantitative explanation.

Second, since the modal contribution to the Local Density of States (LDOS)—or the Purcell factor—is proportional to $Q/\widetilde{V}$ when $Im(\widetilde{V}) \ll Re(\widetilde{V})$ [1], negative mode volumes can reduce the total spontaneous emission rate. Specifically, a quantum emitter located at a point where $Re(\widetilde{V}) < 0$ will experience a dip in its LDOS spectrum, resulting from a negative Purcell factor superimposed on the otherwise flat contribution from non-resonant modes. This should lead to a lowered spontaneous emission.

To validate our prediction, we analyze the system depicted in Fig. 5, excited by a point electric dipole emitter. The emitter is polarized along the z-axis and positioned at the center of the stripe waveguide, where $Re(\tilde{V}^C) < 0$ and $Im(\tilde{V}^C)$ approaches zero, as shown in Fig. 5(d). As illustrated in Fig. S4(a), the total power emitted by the emitter, i.e. the LDOS, exhibits a dip near the resonance frequency, due to the excitation of the QNM with a negative mode volume.

Finally, let us illustrate another intriguing effect of negative mode volumes by introducing gain into the previous computation. If the amount of gain is such that the complex frequency of the fundamental QNM with the smallest decay rate approaches the real axis, it will dominate the response. Moreover, if this mode has a negative mode volume at the location of the dipolar current source, the emitted power will become negative, causing the source to act as a "sink"—extracting energy from the environment rather than supplying it.

To illustrate the "sink" effect, we consider the same system as in Fig. S4(a). To ensure that the QNM with negative mode volume dominates the response, a gain medium is introduced beneath the Fabry-Perot stripe cavity (highlighted in red in Fig. S4(b)). Figure S4(b) depicts the Poynting vector (energy flow) directions in the median plane of the dielectric gap, revealing that energy flows toward the dipole current source, effectively making it a "sink."

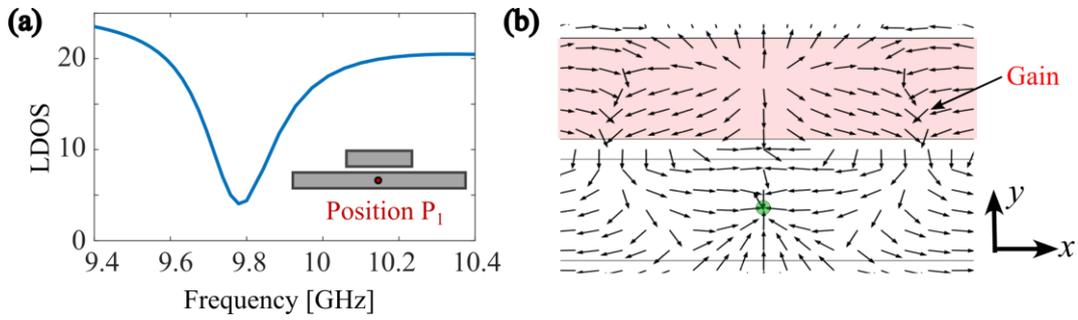

**Figure S4. Unconventional phenomena arising from negative mode volumes.** (**a**) Dip in the normalized LDOS spectrum for a quantum emitter positioned at the center of the waveguide ($P_1$ in the main text, where $Re(\tilde{V}) < 0$) due to the negative Purcell effect. The normalized LDOS is defined as $\gamma/\gamma_0$ with $\gamma = (2/\hbar)\text{Im}[\mathbf{p}^* \cdot \mathbf{E}]$ the LDOS and $\gamma_0 = \omega^3|p|^2/(3\pi\varepsilon_0\hbar c^3)$ the LDOS in vacuum. (**b**) A point-like dipolar current source acts as a tiny sink. The arrows represent the Poynting vector directions in the median plane of the dielectric gap, showing energy flow toward the Dirac source at $P_1$ (green dot). The results are obtained for the geometric parameters of the structure given the caption of Fig. 5. Materials are the same as in the main text for (**a**). In (**b**), a gain material is added beneath the top Fabry-Perot stripe cavity (red region in (**b**)) with permittivity $\varepsilon_{gain} = \varepsilon_{gap} - 0.3i$.